\documentclass[longauth]{aa} 
\usepackage{xcolor}
\usepackage{support-caption}
\usepackage{subcaption}
\usepackage{graphicx}
\usepackage[flushleft]{threeparttable}

\newcommand{\nii}{[\textrm{N}\textsc{ii}]}

\newcommand{\oiii}{[\textrm{O}\textsc{iii}]}
\newcommand{\oii}{[\textrm{O}\textsc{ii}]}
\newcommand{\oiidoub}{[\textrm{O}\textsc{ii}]\ensuremath{\lambda\lambda3727, 3729}}
\newcommand{\oiilam}{[\textrm{O}\textsc{ii}]\ensuremath{\lambda3727}}

\newcommand{\oiiialone}{[\textrm{O}\textsc{iii}]}

\newcommand{\oiiidoub}{[\textrm{O}~\textsc{iii}]\ensuremath{\lambda\lambda4959,5007}}
 
\newcommand{\ha}{\ifmmode {\rm H}\alpha \else H$\alpha$\fi}
\newcommand{\hb}{\ifmmode {\rm H}\beta \else H$\beta$\fi}
\newcommand{\lya}{\ifmmode {\rm Ly}\alpha \else Ly$\alpha$\fi}
\newcommand{\pg}{\ifmmode {\rm P}\gamma \else Pa$\gamma$\fi}
\newcommand{\lyb}{\ifmmode {\rm Ly}\beta \else Ly$\beta$\fi}
\newcommand{\lyg}{\ifmmode {\rm Ly}\gamma \else Ly$\gamma$\fi}

\newcommand{\flyc}{\ifmmode \mathrm{f}_\mathrm{esc}\mathrm{(LyC)} \else $\mathrm{f}_\mathrm{esc}\mathrm{(LyC)}$\fi}

\def\ergs{\ifmmode \mathrm{erg\hspace{1mm}s}^{-1} \else erg s$^{-1}$\fi}

\def\micron{\ifmmode \mu\mathrm{m} \else $\mu$m\fi}
\def\msun{\ifmmode \mathrm{M}_{\odot} \else M$_{\odot}$\fi}
\def\msunyr{\ifmmode \mathrm{M}_{\odot} \hspace{1mm}{\rm yr}^{-1} \else $\mathrm{M}_{\odot}$ yr$^{-1}$\fi}
\def\zsun{\ifmmode Z_{\odot} \else Z$_{\odot}$\fi}
\def\lsun{\ifmmode L_{\odot} \else L$_{\odot}$\fi}
\def\mstar{\ifmmode \mathrm{M}_{\star} \else M$_{\star}$\fi}
\newcommand{\hst}{\textit{HST}}
\newcommand{\jwst}{\textit{JWST}}

\newcommand{\NIRSpec}{\textit{NIRSpec}}
\newcommand{\NIRCam}{\textit{NIRCam}}

\usepackage{txfonts}
\newcommand{\orcid}[1]{\href{https://orcid.org/#1}{\includegraphics[width=10pt]{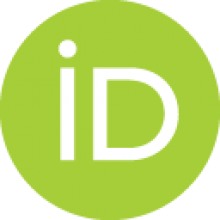}}}
\usepackage{hyperref}
\hypersetup{colorlinks, 
 linkcolor = blue,
 urlcolor = blue,
 citecolor = blue,
 anchorcolor = blue
}
\begin{document} 

\title{New insight on the nature of cosmic reionizers from the CEERS survey}
 
 \subtitle{}
 \author{S. Mascia \orcid{0000-0002-9572-7813}\fnmsep\thanks{E-mail: sara.mascia@inaf.it}
 \inst{1,2}
 \and
 L. Pentericci \orcid{0000-0001-8940-6768}
 \inst{1}
 \and A. Calabrò \orcid{0000-0003-2536-1614}
 \inst{1}
\and P. Santini \orcid{0000-0002-9334-8705}
 \inst{1}
\and L. Napolitano \orcid{0000-0002-8951-4408}
 \inst{1}
 \and P. Arrabal Haro \orcid{0000-0002-7959-8783}
 \inst{3}
 \and M. Castellano \orcid{0000-0001-9875-8263}
 \inst{1}
 \and
 M. Dickinson \orcid{0000-0001-5414-5131}
\inst{3}
 \and P. Ocvirk \orcid{0000-0002-8488-504X} 
 \inst{15}
 \and J. S. W. Lewis \orcid{0000-0001-7917-8474} 
 \inst{19,20}
 \and
 R. Amorín \orcid{0000-0001-5758-1000}
 \inst{17,18}
 \and M. Bagley \orcid{0000-0002-9921-9218}
 \inst{23}
 \and R. Bhatawdekar \orcid{0000-0003-0883-2226}
 \inst{10}
 \and N. J. Cleri \orcid{0000-0001-7151-009X}
 \inst{8,9}
 \and L. Costantin \orcid{0000-0001-6820-0015}
 \inst{4}
 \and A. Dekel \orcid{0000-0003-4174-0374}
 \inst{5}
 \and S. L. Finkelstein \orcid{0000-0001-8519-1130}
 \inst{23}
 \and A. Fontana \orcid{0000-0003-3820-2823}
 \inst{1}
 \and M. Giavalisco \orcid{0000-0002-7831-8751}
 \inst{6}
 \and N. A. Grogin \orcid{0000-0001-9440-8872}
 \inst{13}
 \and N. P. Hathi \orcid{0000-0001-6145-5090}
 \inst{13}
 \and M. Hirschmann \orcid{0000-0002-3301-3321}
 \inst{11,12}
 \and B. W. Holwerda \orcid{0000-0002-4884-6756}
 \inst{16}
 \and I. Jung \orcid{0000-0003-1187-4240}
 \inst{13}
 \and J. S. Kartaltepe \orcid{0000-0001-9187-3605}
 \inst{21}
 \and A. M. Koekemoer \orcid{0000-0002-6610-2048}
 \inst{13}
 \and R. A. Lucas \orcid{0000-0003-1581-7825}
 \inst{13}
 \and C. Papovich \orcid{0000-0001-7503-8482}
 \inst{8, 9}
  \and P. G. Pérez-González \orcid{0000-0003-4528-5639}
 \inst{4}
 \and N. Pirzkal \orcid{0000-0003-3382-5941}
 \inst{22}
 \and J. R. Trump \orcid{0000-0002-1410-0470}
 \inst{14}
  \and S. M. Wilkins \orcid{0000-0003-3903-6935}
 \inst{24,25}
 \and L. Y. A. Yung \orcid{0000-0003-3466-035X}
 \inst{7}
 }
 \institute{\textit{INAF – Osservatorio Astronomico di Roma, via Frascati 33, 00078, Monteporzio Catone, Italy}
 \and 
 \textit{Dipartimento di Fisica, Università di Roma Tor Vergata,
Via della Ricerca Scientifica, 1, 00133, Roma, Italy}
\and
\textit{NSF’s National Optical-Infrared Astronomy Research Laboratory, 950 N. Cherry Ave., Tucson, AZ 85719, USA}
\and
\textit{Centro de Astrobiología (CAB), INTA-CSIC, Ctra de Ajalvir km 4, Torrejón de Ardoz, E-28850, Madrid, Spain}
\and
\textit{Racah Institute of Physics, The Hebrew University of Jerusalem, Jerusalem 91904, Israel}
\and
\textit{University of Massachusetts Amherst, 710 North Pleasant Street,
Amherst, MA 01003-9305, USA}
\and
\textit{Astrophysics Science Division, NASA Goddard Space Flight Center, 8800 Greenbelt Rd, Greenbelt, MD 20771, USA}
\and
\textit{Department of Physics and Astronomy, Texas A\&M University, College Station, TX 77843-4242, USA }
\and
\textit{George P. and Cynthia Woods Mitchell Institute for Fundamental Physics and Astronomy, Texas A\&M University, College Station, TX 77843-4242, USA}
\and
\textit{European Space Agency (ESA), European Space Astronomy Centre (ESAC), Camino Bajo del Castillo s/n, 28692 Villanueva de la Cañada, Madrid, Spain
}
\and 
\textit{Institute for Physics, Laboratory for Galaxy Evolution and Spectral Modelling, EPFL, Observatoire de Sauverny, Chemin Pegasi 51, 1290 Versoix, Switzerland}
\and
\textit{INAF, Osservatorio Astronomico di Trieste, Via Tiepolo 11, 34131 Trieste, Italy}
\and
\textit{Space Telescope Science Institute, 3700 San Martin Drive, Baltimore MD, 21218}
\and
\textit{Department of Physics, University of Connecticut, 196A Auditorium Road, Unit 3046, Storrs, CT 06269 USA}
\and
\textit{Universit\'e de Strasbourg, CNRS, Observatoire astronomique de Strasbourg (ObAS), 11 rue de l’Universit\'e, Strasbourg, France}
\and 
\textit{Physics \& Astronomy Department, University of Louisville, 40292 KY, Louisville, USA}
\and 
\textit{ARAID Foundation. Centro de Estudios de F\'{\i}sica del Cosmos de Arag\'{o}n (CEFCA), Unidad Asociada al CSIC, Plaza San Juan 1, E--44001 Teruel, Spain}
\and
\textit{Departamento de Astronomía, Universidad de La Serena, Av. Juan Cisternas 1200 Norte, La Serena 1720236, Chile}
\and
\textit{Universit\"{a}t Heidelberg, Zentrum f\"{u}r Astronomie, Institut f\"{u}r Theoretische Astrophysik, Albert-Ueberle-Str. 2, 69120 Heidelberg, Germany}
\and
\textit{Max-Planck-Institut f\"{u}r Astronomie, K\"{o}nigstuhl 17, 69117 Heidelberg, Germany}
\and
\textit{Laboratory for Multiwavelength Astrophysics, School of Physics and Astronomy, Rochester Institute of Technology, 84 Lomb Memorial Drive, Rochester, NY, 14623, USA}
\and
\textit{ESA/AURA Space Telescope Science Institute, USA}
\and
\textit{Department of Astronomy, The University of Texas at Austin, Austin, TX, USA}
\and
\textit{Astronomy Centre, University of Sussex, Falmer, Brighton BN1 9QH, UK}
\and
\textit{Institute of Space Sciences and Astronomy, University of Malta, Msida MSD 2080, Malta}
}
 \date{Accepted XXX. Received YYY; in original form ZZZ}

 \abstract
 {The Epoch of Reionization (EoR) began when galaxies grew in abundance and luminosity, so their escaping Lyman continuum (LyC) radiation started ionizing the surrounding neutral intergalactic medium (IGM). Despite significant recent progress, the nature and role of cosmic reionizers are still unclear: in order to define them, it would be necessary to directly measure their LyC escape fraction ($f_{esc}$). However, this is impossible during the EoR due to the opacity of the IGM. Consequently, many efforts at low and intermediate redshift have been made to determine measurable indirect indicators in high-redshift galaxies so that their $f_{esc}$ can be predicted. This work presents the analysis of the indirect indicators of 62 spectroscopically confirmed star-forming galaxies at $6 \leq z \leq 9$ from the Cosmic Evolution Early Release Science (CEERS) survey, combined with 12 sources with public data from other \jwst-ERS campaigns. From the \NIRCam\ and \NIRSpec\ observations, we measured their physical and spectroscopic properties. We discovered that on average $6<z<9$ star-forming galaxies are compact in the rest-frame UV ($r_e \sim $ 0.4 kpc), are blue sources (UV-$\beta$ slope $\sim $ -2.17), and have a predicted $f_{esc}$ of about 0.13. 
 A comparison of our results to models and predictions as well as an estimation of the ionizing budget suggests that low-mass galaxies with UV magnitudes fainter than $M_{1500} = -18$ that we currently do not characterize with \jwst\ observations probably played a key role in the process of reionization.}
\maketitle

\section{Introduction}
The Epoch of Reionization (EoR) is a period in the history of the Universe, occurring roughly during its first billion years, when the hydrogen in the intergalactic medium (IGM) transitioned from a nearly completely neutral to a nearly completely ionized state. This transition was driven by the Lyman continuum (LyC; $\lambda < 912$ \AA) radiation emitted by the first luminous sources that formed in the early Universe. However these sources, i.e. the so-called cosmic reionizers, remain elusive: star-forming galaxies can only account for the photon budget to complete reionization if a substantial fraction of the Ultra-Violet (UV) photons produced by their stellar populations escape from the galaxies' interstellar medium (ISM) and circumgalactic medium (CGM). As a result of the density of star-forming galaxies in the EoR, an average LyC escape fraction ($f_{esc}$) of 10\% across all galaxies is needed \citep[e.g.,][]{Yung2020a, Yung2020b, Finkelstein2019, Robertson2015} to reionize the Universe by $z = 6$, and match the Thomson optical depth of electron scattering in the Cosmic Microwave Background (CMB) \citep{Planck2020}. At $z \geq 4.5$, however, it is impossible to detect the LyC photons escaping from galaxies, since they are absorbed and scattered by the IGM along the line of sight \citep[][]{inoue2014}, and the LyC can only be detected at low and intermediate redshift \citep[e.g.,][]{Flury2022, Izotov2016a, Izotov2016b,Izotov2018a, Izotov2018b, Wang2019,Steidel2018, Fletcher2019, Vanzella2018, Vanzella2020, Marques-Chaves2021, Marques-Chaves2022}. To overcome this problem, key properties of the ISM and conditions that facilitate LyC photons escape (the so-called \textit{indirect indicators}) at lower redshifts have been identified \citep[see][for a review]{Flury2022} and used to infer the average $f_{esc}$ of the cosmic reionizers \citep[e.g.,][]{Jung2023, Mascia2023, Roy2023, Saxena2023b}. 

The relative importance of massive and low-mass galaxies in driving reionization is still a matter of great debate as it is intrinsically related to the timeline and topology of reionization. It is expected that reionization starts earlier, and perhaps proceeds in a spatially more homogeneous manner, when faint and low-mass galaxies with a higher $f_{esc}$ dominate ionizing photon budgets over bright galaxies \citep[e.g.,][]{Ferrara2013, Finkelstein2019, Dayal2020}. Conversely, a relatively delayed reionization process is predicted when the contributions from faint galaxies ($M_{1500} \geq -18$) are subdominant to that from brighter systems \citep{Robertson2015, Naidu2020}. While both types of galaxies are likely to contribute to the ionizing budget, the balance and interplay between them remain uncertain. 
\\
To understand the role of faint and bright sources, we need to determine what is their relative contribution to   the total ionizing emissivity ($\dot{n}_{ion}$), i.e., the number of ionizing photons emitted per unit time and comoving volume \citep[see][for a detailed review]{Robertson2022} which is commonly expressed as:
\begin{equation}
 \dot{n}_{ion} = f_{esc} \ \xi_{ion} \ \rho_{UV},
 \label{eq:N_ion}
\end{equation}

\noindent in which $\xi_{ion}$ is the ionizing photon production efficiency, i.e. the number of produced ionizing photons per UV luminosity density,  the $\rho_{UV}$ is the integral of the UV luminosity function (the number of galaxies per UV luminosity per comoving volume), and $f_{esc}$ is the fraction of ionizing photons that reaches the IGM.
\\
In the above equation, the $\rho_{UV}$ of galaxies is relatively well-constrained up to the very high-redshift Universe \citep[e.g.][]{Bouwens2015, Bouwens2021, Donnan2022}.  We know that many factors influence the photon production efficiency, including the initial mass function, the stellar metallicity, the evolution of individual stars, and possible stellar binary interactions \citep[e.g.,][]{Zackrisson2011, Zackrisson2013, Zackrisson2017, Eldridge_2017, Stanway2018, Stanway2019}.  A commonly accepted value is  $\log\xi_{ion}=25.3$ but   many recent observations at intermediate and high redshifts \citep[e.g.,][]{Matthee2017, Izotov2017, Nakajima2018b, Shivaei2018, Lam2019, Bouwens2016, Stark_2015, Stark2017, Atek2022, Castellano2022,Castellano2023}. \cite{Yung2020b} demonstrated that $\xi_{ion}$ can vary quite widely as a function of galaxy properties, and a fixed value is just not sufficient to properly capture the scatter in a large population of galaxies. With the James Webb Space Telescope (\jwst), we are now able to measure $\xi_{ion}$ from the rest-frame optical lines \citep[e.g.,][]{Schaerer2016, Shivaei2018, Chevallard2013, Tang2019}, instead of adapting the same value for the entire galaxy population. 
The only remaining big uncertainty in the emissivity equation is thus the escape fraction and how it varies with $M_{1500}$ (or stellar mass), which is the subject of this work.

In \cite{Mascia2023} (M23 hereafter), we have  shown that at the end of reionization ($4.5 \leq z \leq 8$), star-forming galaxies are often compact ($r_e \simeq 0.2-0.5$ kpc), and with blue UV slopes (median $\beta = -2.08$). Moreover, the analyzed sources present properties (in terms of the \oiiidoub/\oiilam\ line ratios, O32 hereafter, H$\beta$ rest-frame equivalent widths, $EW_0(\hb)$, UV-$\beta$ slopes, $r_e$, and $\Sigma_{SFR}$) consistent with those of low-z galaxies with measured $f_{esc}$ larger than 0.05.  These results  suggested that the average low mass galaxies around the EoR have physical and spectroscopic properties consistent with moderate escape of ionizing photons ($f_{esc}=0.1-0.2$), resulting in a dominance of low-mass, faint galaxies during cosmic reionization. The results of M23 may clarify the role of faint galaxies during reionization, but were based on a very limited sample of sources. 
In this work we use the \jwst/Near InfraRed Spectrograph (\NIRSpec) and Near InfraRed Camera (\NIRCam) observations from the Cosmic Evolution Early Release Science (CEERS) survey of a much larger sample of high redshift galaxies to  probe their role as cosmic reionizers during the EoR and put the conclusions of M23 on firmer grounds. 

This paper is organized as follows: we present the data set in Sec.~\ref{sec:data}. We characterize the selected sample in Sec.~\ref{sec:meas}, and compare the physical and spectroscopic properties with models in Sec.~\ref{sec:analysis}. In Sec.~\ref{sec:discussion} we estimate the total ionizing budget from our sample and discuss our results, while in Sec.~\ref{sec:conclusions} we summarize our key conclusions. Throughout this work, we assume a flat $\Lambda$CDM cosmology with $H_0$ = 67.7 km s$^{-1}$ Mpc$^{-1}$ and $\Omega_m$ = 0.307 \citep{Planck2020} and the \cite{Chabrier2003} initial mass function. All magnitudes are expressed in the AB system \citep{Oke1983}.

\section{Data}\label{sec:data}
\subsection{CEERS-JWST data}

We used \jwst/\NIRSpec\ observations from the Cosmic Evolution Early Release Science survey (CEERS; ERS 1345, PI: S. Finkelstein) in the CANDELS Extended Groth Strip (EGS) field \citep[][]{Grogin2011, Koekemoer2011}. The final list of targets selected for spectroscopic observations during the CEERS program and the way in which targets have been prioritized will be presented in Finkelstein et al. \citep[in prep, see also][]{Finkelstein2022, Finkelstein2022b}, while the \NIRSpec\ data will be described in Arrabal Haro et al. (in prep.), see also \cite{ArrabalHaro2023}. 
We also use the CEERS \NIRCam\ imaging in six broadband filters (F115W, F150W, F200W, F277W, F356W and F444W) and one medium-band filter (F410M) over 10 pointings. Details on imaging data reduction and analysis are presented in \cite{Bagley2023} \citep[see also][]{Finkelstein2022, Finkelstein2022b}.

In this section, we provide a brief summary, highlighting the most relevant points and explaining the methods we used to study the properties of the galaxies of our sample.

The focus of this study is on all sources at $6\leq z \leq 9$. We selected all the sources with a photometric redshift higher than 5 that have a \NIRSpec\ spectrum obtained either with the three medium-resolution ($R \approx 1000$) grating spectral configurations (G140M/F100LP, G235M/F170LP and G395M/F290LP), which, together, cover wavelengths between 0.7-5.1 $\mu$m,  or with the PRISM/CLEAR configuration, which provides continuous wavelength coverage of 0.6-5.3 $\mu$m with spectral resolution ranging from $R \sim 30$ to 300. We  visually examined all these spectra for detectable optical lines and measured the systemic redshifts of 70 sources in the chosen range, using the \hb, \oiiidoub, and (when present) \ha\ lines. The best redshift solution was determined by fitting single Gaussian functions to the strongest emission lines and combining the centroids of the fits. In 66 cases, the \oiidoub, \oiii\ and/or \hb\ were detected and their line fluxes were measured. For the remaining 4 cases, the redshifts were obtained by fitting the \ha\ line alone, so they are formally included in our sample but they can not be used for further analysis since this is the only line present in the spectra. For this part of our analysis, we use \textsc{Mpfit}\footnote{\url{http://purl.com/net/mpfit}} \citep{markwardt2009}. 
Note that with the PRISM's resolution of $R > 140$ at $\lambda > 3.4\,\mu$m, we are able to discern \hb\ from \oiiialone, and  resolve the \oiiialone\ doublet but we do not resolve  the \ha\ + \nii\ doublet.

All CEERS MSA IDs, coordinates, and spectroscopic redshifts are reported in Table \ref{tab:summary1} along with their spectroscopic and physical properties, whose determination is described in the next sections.
Some of the sources presented in this work have been already identified and analyzed in previous works, specifically \cite{Jung2023} (MSA IDs: 686, 689, 698), \cite{Fujimoto2023} (MSA IDs: 2, 3, 4, 7, 20, 23, 24), \cite{ArrabalHaro2023} (MSA IDs: 80025, 80083), \cite{Larson2023} (MSA ID:1019), and \cite{Tang2023} (MSA IDs: 3, 23, 24, 44, 407, 498, 499, 686, 689, 698, 717, 1019, 1023, 1025, 1027, 1029, 1038, 1102, 1143, 1149, 1163). 

\begin{figure}
\centering
\includegraphics[width=\linewidth]{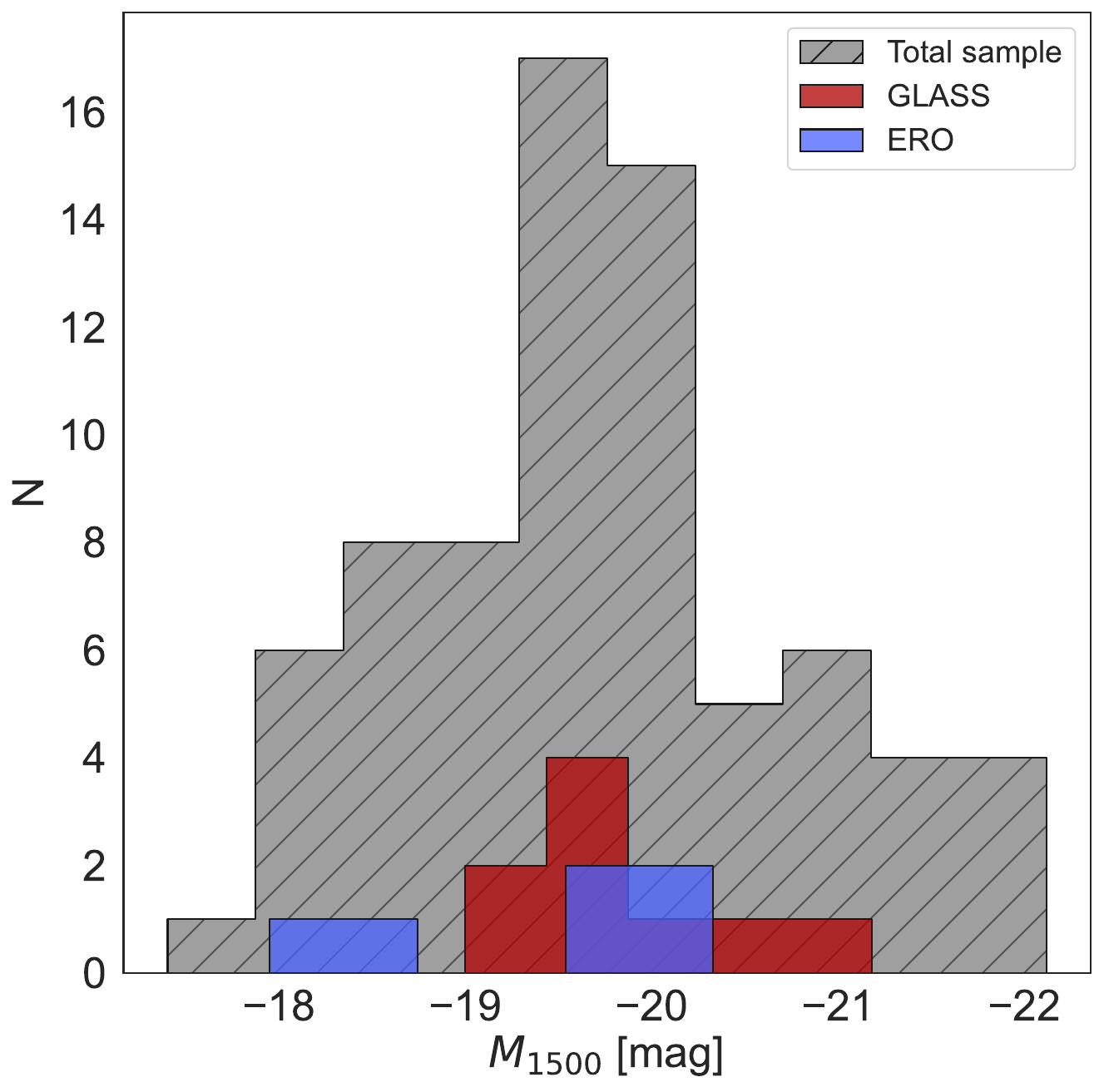}
\caption{$M_{1500}$ distribution for the analyzed sources at $6 \leq z \leq9$ \citep[grey: total sample; red and blue are respectively the GLASS sample and the ERO sample from][]{Mascia2023, Noirot2022}.
\label{fig:mag}}
\end{figure}

\begin{table*}
\caption{Physical and spectroscopic properties  for the CEERS sample.}\label{tab:summary1}
\tiny{
$$ 
\begin{array}{llcccllccc}
\hline \hline
\noalign{\smallskip}
\text{MSA ID} & \text{RA} \ [\text{deg}] & \text{DEC} \ [\text{deg}] & z_{spec} & \beta &EW_0(\hb) \ [\AA]& \text{O32} & r_e \ [\text{kpc}] & f_{esc} \text{ (pred.)} & \log \xi_{ion}\\
\noalign{\smallskip}
 \hline
 \noalign{\smallskip}
 2^* & 214.994402 & 52.989379 & 8.803 & -1.55 \pm 0.09 & <178 & 3.7 \pm 0.3 & < 0.12 & > 0.07 & 25.40 \pm 0.94\\
3^* & 215.005189 & 52.996580 & 8.007 & -2.63 \pm 0.67 & 133 \pm 81 & 10.3 \pm 0.2 & 0.50 \pm 0.18 & 0.20 \pm 0.17 & 26.04 \pm 1.02\\
4^* & 215.005365 & 52.996697 & 7.995 & -2.07 \pm 0.17 & 144\pm 121& -& < 0.12& 0.18 \pm 0.16 & 25.49 \pm 0.95\\
7^* & 215.011706 & 52.988303 & 8.871 & -2.24 \pm 0.12 & 296 \pm 55 & 5.3 \pm 2.1 & 0.22 \pm 0.02 & 0.15 \pm 0.11 & 25.12 \pm 0.90\\
20^* & 214.830685 & 52.887771 & 7.764 & -1.25 \pm 0.31 & 57\pm 13& 2.2 \pm 0.4 & 0.20 \pm 0.03 & 0.03 \pm 0.02 & 24.38 \pm 0.80\\
23^*& 214.901252 & 52.846997 & 8.883 & -1.42 \pm 0.58 & 187 \pm 83 & 4.2 \pm 0.3 & 0.12 \pm 0.03 & 0.07 \pm 0.06 & 26.02 \pm 1.02\\
24^* & 214.897232 & 52.843854 & 9.000& -1.88 \pm 0.69 & -& > 8& < 0.12 & > 0.06 & 25.68 \pm 0.98\\
44^* & 215.001115 & 53.011269 & 7.106 & -2.58 \pm 0.12 & 139\pm 11& 18.4 \pm 4.5 & < 0.13 & > 0.39 & 25.45 \pm 0.94\\
67^* & 215.015597 & 53.011857 & 6.205 & -2.89 \pm 0.52 & 21\pm 10& 0.2 \pm 0.1& 0.15 \pm 0.07 & 0.07 \pm 0.06 & 24.56 \pm 0.82\\
355^* & 214.806482 & 52.878827 & 6.102 & -2.05 \pm 0.09 & 78\pm 18& > 23& 0.46 \pm 0.06 & 0.05 \pm 0.03 & 25.29 \pm 0.92\\
362^* & 214.812689 & 52.881536 & 6.052 & -2.59 \pm 0.23 & 66\pm 33& > 29& 0.43 \pm 0.14 & 0.10 \pm 0.08 & 25.60 \pm 0.96\\
386^* & 214.832184 & 52.885083 & 6.615 & -1.82 \pm 0.24 & 132\pm 47& 1.4 \pm 0.1 & 0.42 \pm 0.12 & 0.03 \pm 0.02 & 25.99 \pm 1.02\\
390^* & 214.811038 & 52.868521 & 6.295 & -2.13 \pm 0.07 & 303 \pm 40 & 123.0 \pm 12.6 & 0.76 \pm 0.20 & 0.24 \pm 0.20 & 25.76 \pm 0.99\\
397^* & 214.836197 & 52.882693 & 6.003 & -2.17 \pm 0.03 & 487\pm 53& 12.8 \pm 0.5 & 0.46 \pm 0.02 & 0.14 \pm 0.11 & 25.97 \pm 1.02\\
407^* & 214.839316 & 52.882565 & 7.031 & -2.20 \pm 0.21 & 108\pm 17& 1.1 \pm 0.1 & 0.17 \pm 0.04 & 0.07 \pm 0.05 & 25.68 \pm 0.97\\
428^* & 214.824551 & 52.868856 & 6.104 & -2.09 \pm 0.03 & -& - & < 0.15 & - & -\\
439^* & 214.825364 & 52.863065 & 7.181 & -2.60 \pm 0.15 & 60\pm 28& -& 0.15 \pm 0.03 & 0.17 \pm 0.12 & 25.38 \pm 0.93\\
476^* & 214.805561 & 52.836345 & 6.017 & -2.05 \pm 0.09 & 17\pm 3& > 4& < 0.15 & > 0.07 & 24.13 \pm 0.76\\
481^* & 214.827785 & 52.850615 & 6.932 & -2.15 \pm 0.06 & < 79 & 2.1 \pm 0.2 & 0.14 \pm 0.03 & 0.10 \pm 0.07 & 24.30 \pm 0.78\\
496^* & 214.864735 & 52.871719 & 6.571 & -2.28 \pm 0.13 & 25\pm 9& -& 0.24 \pm 0.04 & 0.02 \pm 0.01 & 24.80 \pm 0.85\\
498^* & 214.813045 & 52.834249 & 7.180 & -2.50 \pm 0.07 & 446\pm 23& 9.6 \pm 2.2 & 0.30 \pm 0.02 & 0.23 \pm 0.17 & 24.62 \pm 0.83\\
499^* & 214.813004 & 52.834170 & 7.171 & -1.83 \pm 0.38 & 76\pm 24& 4.5 \pm 1.5 & 0.25 \pm 0.01 & 0.08 \pm 0.06 & 25.14 \pm 0.90\\
535^*& 214.859175 & 52.853587 & 7.117 & -2.08 \pm 0.01 & 130\pm 25& -& -& -& 25.82 \pm 0.99\\
542^* & 214.831624 & 52.831505 & 7.061 & -2.43 \pm 0.18 & 114 \pm 13 & -& 0.15 \pm 0.03 & 0.19 \pm 0.13 & -\\
568^* & 214.891863 & 52.869054 & 6.806 & -2.09 \pm 0.11 & -& - & 0.50 \pm 0.28 & - & -\\
577^* & 214.892861 & 52.865157 & 6.703 &-2.07 \pm 0.13 & 370 \pm 41 & - & < 0.14 & > 0.37 &25.52 \pm 0.95\\
603^* & 214.867247 & 52.836737 & 6.059 & -2.17 \pm 0.10 & 166\pm 87& 2.0\pm 0.1& 2.05 \pm 0.06 & 0.01 \pm 0.01 & 26.00 \pm 1.02\\
613^* & 214.882077 & 52.844346 & 6.731 & -1.98 \pm 0.06 & 132 \pm 38 & 0.5 \pm 0.1& 0.26 \pm 0.04 & 0.03 \pm 0.02 & 25.60 \pm 0.96\\
618^* & 214.876469 & 52.839412 & 6.050 & -2.16 \pm 0.12 & 178\pm 23& 0.6 \pm 0.1& 0.79 \pm 0.12 & 0.02 \pm 0.01 & 24.96 \pm 0.88\\
648^* & 214.899823 & 52.847647 & 6.054 & -2.05 \pm 0.09 & 15\pm 11& 0.3 \pm 0.1& 0.28 \pm 0.13 & 0.03 \pm 0.02 & 24.51 \pm 0.81\\
686 & 215.150862 & 52.989562 & 7.754 & -3.69 \pm 0.89 & 123\pm 4& -& 0.25 \pm 0.06 &0.59 \pm 0.48 & 25.27 \pm 0.92\\
689 & 214.999052 & 52.941977 & 7.548 & -1.43 \pm 0.65 & 137\pm 61& 8.3 \pm 1.0& 0.43 \pm 0.10 & 0.06 \pm 0.05 & 26.05 \pm 1.03\\
698 & 215.050317 & 53.007441 & 7.473 & -1.72 \pm 0.29 & 134\pm 5& 20.0 \pm 1.8 & 0.38 \pm 0.03 & 0.13 \pm 0.10 & 25.48 \pm 0.95\\
716 & 215.080349 & 52.993241 & 6.964 & -& -& - & -& - & -\\
717 & 215.081406 & 52.972180 & 6.933 & -1.75 \pm 0.35 & -& 5.4 \pm 0.8 & 0.77 \pm 0.08 & 0.03 \pm 0.03 & 25.16 \pm 0.90\\
749^* & 215.002840 & 53.007588 & 7.090 & -1.82 \pm 0.05 & 142\pm 12& > 6& < 0.13 & > 0.12 & 24.27 \pm 0.78\\
792^* & 214.871766 & 52.833167 & 6.259 & -1.73 \pm 0.25 & 43\pm 12& 3.3 \pm 0.3 & 1.01 \pm 0.26 & 0.02 \pm 0.01 & 25.65 \pm 0.97\\
829^* & 214.861594 & 52.876159 & 7.168 & -2.05 \pm 0.20 & 39\pm 19& 3.8 \pm 0.2 & 0.35 \pm 0.07 & 0.08 \pm 0.06 & 25.51 \pm 0.95\\
\textcolor{magenta}{1019^*} & 215.035391 & 52.890662 & 8.680 & -2.22 \pm 0.06 & 63\pm 10& 12.7 \pm 0.9 & 0.52 \pm 0.02 & 0.13 \pm 0.10 & 27.02 \pm 1.16 \\
1023 & 215.188413 & 53.033647 & 7.778 & -2.20 \pm 0.97 & 49\pm 18& 1.0\pm 0.4 & 0.80 \pm 0.13 & 0.03 \pm 0.03 & 24.71 \pm 0.84\\
1025^* & 214.967547 & 52.932953 & 8.716 & -2.18 \pm 0.10 & 456\pm 44& 7.6 \pm 1.2 & < 0.12 & 0.22 \pm 0.16 & 26.25 \pm 1.05\\
1027^* & 214.882994 & 52.840416 & 7.822 & -1.71 \pm 0.07 & 127\pm 28& 13.2 \pm 2.3 & < 0.13 & 0.17 \pm 0.12 & 25.44 \pm 0.94\\
1029 & 215.218762 & 53.069862 & 8.613 & -& 32\pm 4& 3.4 \pm 0.9 & 0.73 \pm 0.08 & - & 24.46 \pm 0.81\\
1038^* & 215.039697 & 52.901597 & 7.196 & -1.62 \pm 0.11 & 105\pm 63& 3.0 \pm 0.1 & 0.41 \pm 0.08 & 0.04 \pm 0.03 & 25.16 \pm 0.90\\
1064 & 215.177167 & 53.048975 & 6.802 & -3.27 \pm 0.65 & 26\pm 3& -& 0.24 \pm 0.04 & 0.24 \pm 0.21 & 24.83 \pm 0.86\\
1065 & 215.116854 & 53.001081 & 6.192 & -2.18 \pm 0.94 & 24\pm 8& 2.7 \pm 0.6 & 0.51 \pm 0.15 & 0.07 \pm 0.07 & 24.76 \pm 0.85\\
1102 & 215.091047 & 52.954285 & 6.998 & -2.57 \pm 0.90 & -& 6.8 \pm 1.3 & 0.61 \pm 0.18 & 0.13 \pm 0.13 & 24.64 \pm 0.83\\
1115^* & 215.162818 & 53.073097 & 6.302 & -& 93\pm 6& 2.0\pm 0.1& < 0.11 & - & 25.97 \pm 1.02\\
1142 & 215.060716 & 52.958708 & 6.962 & -1.56 \pm 0.44 & -& 17.7 \pm 5.2 & 0.46 \pm 0.12 & 0.09 \pm 0.07 & -\\
1143 & 215.077006 & 52.969504 & 6.93 & -2.89 \pm 0.67 & 27\pm 1& 5.6 \pm 0.2 & 0.30 \pm 0.07 & 0.25 \pm 0.20 & 25.03 \pm 0.89\\
1149 & 215.089714 & 52.966183 & 8.177 & -1.50 \pm 0.62 & 291\pm 21& 7.4 \pm 0.6 & 0.38 \pm 0.09 & 0.06 \pm 0.05 & 25.66 \pm 0.97\\
1160 & 214.805047 & 52.845877 & 6.569 & -2.20 \pm 0.97 & 93\pm 27& > 18& 0.26 \pm 0.08 & 0.13 \pm 0.13 & 25.51 \pm 0.95\\
1163 & 214.990468 & 52.971990 & 7.450 & -3.12 \pm 0.76 & 19\pm 11& 0.6 \pm 0.1& 0.61 \pm 0.16 & 0.07 \pm 0.07 & 24.97 \pm 0.88\\
1414^* & 215.128029 & 52.984936 & 6.678 & -2.01 \pm 0.02 & -& - & -& - & -\\
1518 & 215.006802 & 52.965041 & 6.110 & -2.95 \pm 0.51 & 124\pm 87& 24\pm 4.8 &0.64 \pm 0.08 & 0.28 \pm 0.22 & 25.62 \pm 0.97\\
1558^* & 214.830637 & 52.835297 & 6.884 & -2.05 \pm 0.01 & 17 \pm 2 & 0.4 \pm 0.1 & < 0.14 & > 0.57 & 23.97 \pm 0.74\\
1561 & 215.166097 & 53.070755 & 6.198 & -3.49 \pm 0.67 & 95\pm 6& 15.1 \pm 3.2 & 0.48 \pm 0.07 & 0.38 \pm 0.24 & 25.44 \pm 0.94\\
2355^* & 215.008489 & 52.977973 & 6.112 & -2.06 \pm 0.04 & 49\pm 7& 3.4 \pm 0.6 & 0.18 \pm 0.03 & 0.10 \pm 0.07 & 24.83 \pm 0.86\\
23642 & 215.230033 & 53.015572 & 6.909 & -2.82 \pm 0.40 & -& 10.1 \pm 1.3 & 1.30 \pm 0.23 & 0.13 \pm 0.13& -\\
28944^* & 214.867500 & 52.836872 & 6.056 & -1.73 \pm 0.05 & 24 \pm 12 & 27.2 \pm 5.6 & 2.11 \pm 0.06 & 0.02 \pm 0.01 & 24.82 \pm 0.85\\
31329 & 215.055116 & 53.000850 & 6.144 & -1.89 \pm 0.37 & 46\pm 7& 9.5 \pm 0.6 & 1.26 \pm 0.21 & 0.03 \pm 0.02 & 24.89 \pm 0.87\\
80025^* & 214.806065 & 52.750867 & 7.655 & -1.90 \pm 0.18 & 83\pm 9& 8.2 \pm 1.6 & 0.44 \pm 0.07 & 0.08 \pm 0.06 & 25.55 \pm 0.96\\
80083^* & 214.961276 & 52.842364 & 8.635 & -1.59 \pm 0.07 & 75\pm 14& 28\pm 1.6 & 0.24 \pm 0.06& 0.18 \pm 0.13 & 25.34 \pm 0.93\\
80710^* & 214.884985 & 52.836045 & 6.552 & -3.07 \pm 0.38 & 47 \pm 8 & - & < 0.14 & > 0.27 & 24.67 \pm 0.84\\
80917^* & 214.933838 & 52.845785 & 6.155 & -2.18 \pm 0.05 & 39 \pm 9 & 0.9 \pm 0.1 & 0.65 \pm 0.19 &0.03 \pm 0.02 & 24.87 \pm 0.86\\
80925^* & 214.948680 & 52.853273 & 6.754 &-1.97 \pm 0.05 & 230 \pm 26 & 4.9 \pm 0.5 & 0.65 \pm 0.12 & 0.05 \pm 0.03 & 26.08 \pm 1.03\\
80374^* & 214.898074 & 52.824895 & 7.178 & -2.25 \pm 0.01 & 321 \pm 119 & > 16& 0.26 \pm 0.12& 0.24 \pm 0.18 & 25.87 \pm 1.00\\
80596^* & 214.771865 & 52.778189 & 6.544 &-2.00 \pm 0.05 & 37 \pm 17 & 15.9 \pm 4.6 & 0.74 \pm 0.19 & 0.08 \pm 0.07 & 25.00 \pm 0.88\\
81063^* & 214.799110 & 52.725119 & 6.094 & -2.00 \pm 0.05& 59 \pm 23 & 31.0 \pm 3.7 & 0.49 \pm 0.07 & 0.18 \pm 0.14 & 25.72 \pm 0.98\\
81068^* & 214.820507 & 52.737148 & 6.276 &-2.01 \pm 0.05 & 74 \pm 3 & 18.1 \pm 6.5 & < 0.14 & > 0.26 & 25.17 \pm 0.90\\
\noalign{\smallskip}
\hline
\end{array}
$$
\begin{tablenotes}
 \item \small $^*$: NIRCam photometry available. $r_e$ with errors: determined in F150W (or F200W for ID:542) with \textsc{Galight}. Due to its nature as an AGN, ID:1019 (in magenta) is excluded from the final sample \citep{Larson2023}.
 \end{tablenotes}}
\end{table*}

\subsection{Data from other programs}

Several additional public sources are used to expand our EoR sample. 
In M23, we examined a sample of sources observed from the GLASS-ERS program (PID 1324, PI: T. Treu) using three high-resolution (R $\sim$ 2000-3000) spectral configurations (G140H/F100LP, G235H/F170LP, and G395H/F290LP). For the purpose of this work  we specifically selected the 7 sources at $z_{spec} > 6$ (GLASS-JWST IDs: 10000, 10021, 100001, 100003, 100005, 150008, 400009), along with 2 additional sources at $z_{spec} > 6$ from a DDT program (PID 2756, PI: W. Chen), which were obtained using the PRISM/CLEAR configuration (DDT IDs: 10025, 100004). All these sources are located in the Abell 2744 cluster field.

From the spectroscopic redshift catalogue by \cite{Noirot2022}, we selected 4 more sources from the Early Release Observations (ERO) program on the galaxy cluster SMACS J0723.3-7327 at $z_{spec} > 6$ (ERO IDs: 4590, 5144, 6355, 10612).  These spectra were acquired with medium resolution spectral configurations (G235M and
G395M). The properties we use in this work were derived from \cite{Trussler2022} and \cite{Schaerer_2022}. For all the above sources, IDs, coordinates, spectroscopic redshifts, spectroscopic, and physical properties are reported in Appendix 1, Table \ref{tab:summary}.

\section{Method}\label{sec:meas}

\subsection{Measurements of physical parameters}

 We measured the physical parameters of the CEERS sample as described in \cite{Santini2022}, 
 by fitting synthetic stellar templates with \textsc{zphot} \citep{fontana00} to the seven-band \NIRCam\ photometry \citep[][for the sources marked with $^*$ in Table \ref{tab:summary1}]{Finkelstein2023} and the released HST photometry \citep{Stefanon2017}. Specifically we measured 
 the stellar masses $M_{\star, obs}$, the observed absolute UV magnitudes at 1500\AA\ ($M_{1500, obs}$), the dust reddening $E(B-V)$ and the ages. We adopted \cite{Bruzual2003} models and assumed delayed exponentially declining star formation histories -- SFH($t$)$\propto (t^2/\tau) \cdot \exp(-t/\tau)$ -- with $\tau$ ranging from 0.1 to 7 Gyr. The age ranges from 10 Myr to the age of the Universe at each galaxy redshift, while metallicity can assume values of 0.02, 0.2 or 1 times Solar metallicity. For the dust extinction, we used the \cite{Calzetti2000} law with $E(B-V)$ which can assume values ranging from 0, 0.03, 0.06, 0.1, 0.15, and from 0.2 to 1.1 in step of 0.1. We computed $1\sigma$ uncertainties on the physical parameters by retaining, for each object, the minimum and maximum fitted masses among all the solutions with a probability $P(\chi^2)>32\%$ of being correct, fixing the redshift to the best-fit value. In Fig. \ref{fig:mag} we present the $M_{1500, obs}$ distribution of the CEERS sources in our sample, which ranges from $-22$ to $-18$ AB mag. For reference, we also show the distribution of the $M_{1500}$ for the GLASS and ERO sources we are considering in this work.

\subsection{Dust correction and emission line flux measurements} \label{sec:data_line}

We measured the total flux of each detected line (Balmer lines, \oii, and \oiiialone) with a single Gaussian fit. From the flux measurement we subtracted a constant continuum emission, which is estimated from a wavelength region adjacent ($\pm 160 \AA$) to the emission line. When the continuum was not well constrained (signal-to-noise ratio S/N $<2$) from the fit, we estimated it subtracting the line contribution to the F444W photometry, following \cite{Fujimoto2023}. When the S/N of \oii, \oiiialone, or \hb\ was less than 2, we set $2\sigma$ as an upper limit. 

Prior to carrying out a quantitative analysis, it is necessary to consider corrections for dust reddening.
For 28 galaxies, \ha\ and \hb\ are both available and we calculated the correction for dust extinction on the basis of the Balmer decrement, assuming a \cite{Calzetti2000} extinction law and an intrinsic ratio \ha/\hb\ = 2.86 \citep[see e.g.,][]{Dominguez2013, Kashino2013, Price2014}, which is valid for an electron temperature of 10000 K. The nebular $E(B-V)$ determined from the Balmer decrement are in agreement with the stellar reddening determined from the SED fitting. Therefore for the 38 sources in the sample without \ha, 
we converted their stellar $E(B-V)_{SED}$ to nebular $E(B-V)$ following \cite{Calzetti2000} and applied the nebular corrections derived from these values. 

With the dust corrected spectra, we calculated the O32 line ratios and the \oiiialone\ and/or \hb\ rest-frame $EW$s. We list all these values in Table \ref{tab:summary1}. Within the errors, our measurements are consistent with those from previous works for sources in common \citep{Jung2023,Fujimoto2023,ArrabalHaro2023, Tang2023}.

\subsection{UV-$\beta$ slopes}
We measured the UV-$\beta$ slope of our galaxies from the \NIRCam\ photometry and/or the previously available \hst\ photometry \citep{Stefanon2017}, with the approach detailed in \cite{Calabro2021}.
We considered all the photometric bands whose entire bandwidths are between $1216$ and $3000$ \AA\ rest frame. The former limit is set to exclude the \lya\ line and Ly-break, while the latter limit is slightly larger than that adopted in \cite{Calabro2021} to ensure that we can use more bands. 

We then fitted the selected photometry with a single power-law of the form f($\lambda$) $\propto \lambda^\beta$ \citep{Calzetti1994, Meurer1999}. 
In practice, we fitted the available photometric bands amongst \hst\ F125W, F140W, F160W and \jwst-\NIRCam\ F115W, F150W or F200W depending on the exact redshift of the sources. This choice allows us to uniformly probe the spectral range between $1500$ and $3000$ \AA\ for most of the galaxies. 
We measured the $\beta$ and associated uncertainty for each source using a bootstrap method: by using $n = 1000$ Monte Carlo simulations, the fluxes in each band were resampled according to their error. The results provided a slope distribution from which we calculated the mean and standard deviation of $\beta$ for each galaxy. Two of the sources in our sample did not have the necessary data, so we were able to estimate the $\beta$ slopes only for 64 galaxies.
The results on $\beta$ with associated errors are reported in Table \ref{tab:summary1}.
We note that for 5 sources different $\beta$ slopes are published in literature \citep[][]{Jung2023, ArrabalHaro2023}, but they were estimated from SED fitting or from the spectra. For 4 out of 5, our values are consistent with the published ones within the uncertainties.

\begin{figure}
\includegraphics[width=\linewidth]{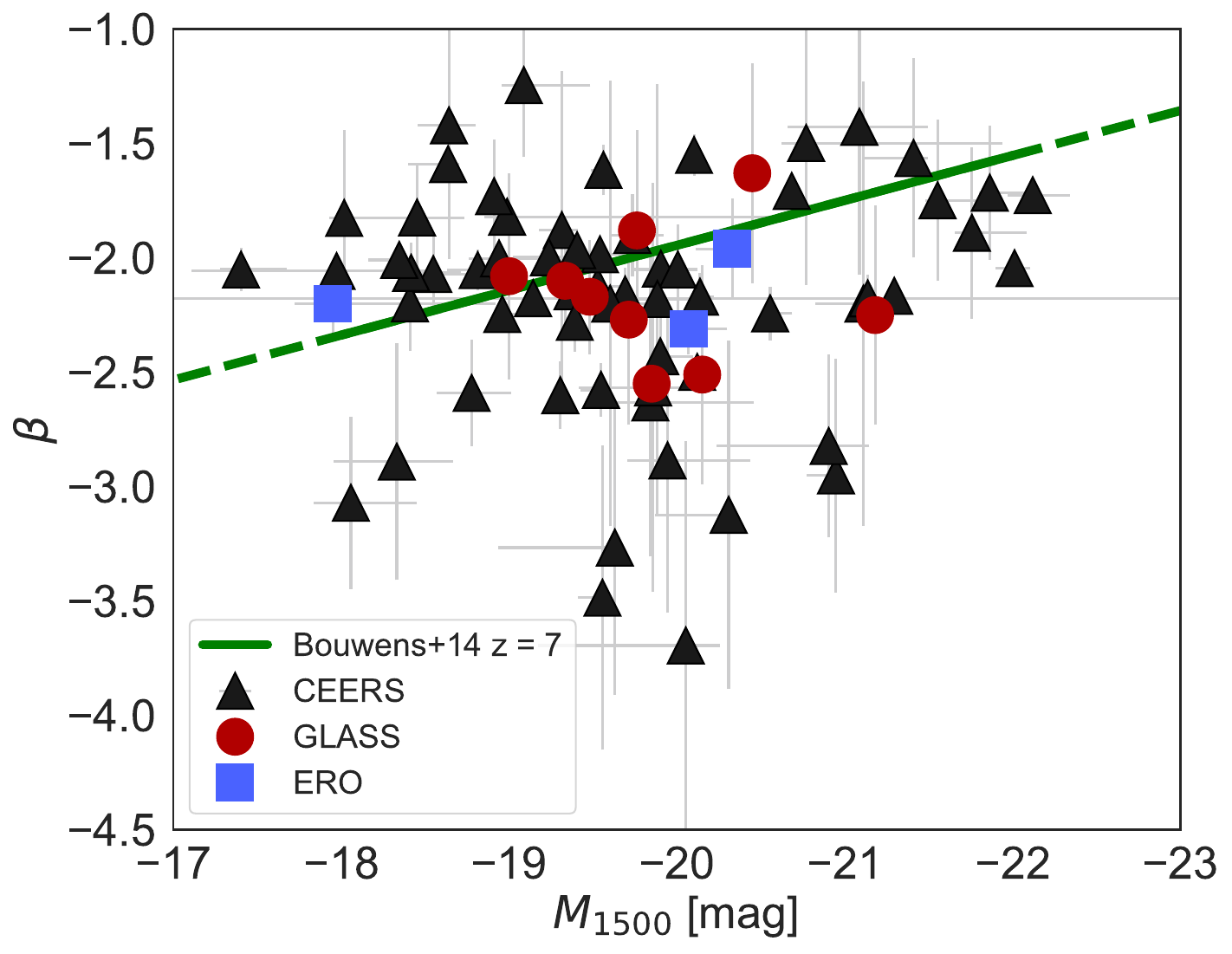}
\caption{$\beta$ vs. $M_{1500}$. Black triangles: CEERS sources with $\beta$ slope obtained fitting 3 or 2 photometric bands. Red dots: GLASS sample; blue squares: ERO sample. The green line shows the relation at $z \sim 7$ derived from HST data by \cite{Bouwens2014}. Dashed portions indicate the extrapolation of the relation in our range of $M_{1500}$. 
 \label{fig:beta_muv}}
\end{figure}

In Fig. \ref{fig:beta_muv} we show the relation between our measured $\beta$ slope values and $M_{1500}$ and the observed trend at $z = 7$ from \cite{Bouwens2014}. We also plot the $\beta$ values as function of $M_{1500}$ for the GLASS and ERO sample. Our results are consistent with the best fit relation from \cite{Bouwens2014} although with a large scatter. We must notice that the galaxies with the bluer slopes (with values around -3) that most deviate from the relation also have the largest uncertainties. Overall we confirm the existence of a broad correlation between $\beta$ and UV magnitude at $z \sim 7$ \citep[e.g.,][]{Wilkins2011, Finkelstein2012, Bouwens2014, Nanayakkara2022}. Our average $\beta$ value at $z \sim 7$, $\langle \beta\rangle = -2.17 \pm 0.47$, is in good agreement with \cite{Dunlop2013} ($\langle \beta\rangle = -2.1 \pm 0.2$ at $z \sim 7$).

\subsection{UV half-light radii}

We measure the half-light radius $r_e$ of each galaxy in the rest-frame UV using the python software \textsc{Galight}\footnote{\url{ https://github.com/dartoon/galight}} \citep{Ding2020}, which adopts a forward-modeling technique to fit a model to the observed luminosity profile of a source. 
We assume that the galaxies are well represented by a Sersic profile \citep{sersicpaper}. In the fitting process, we constrain the axial ratio $q$ to the range $0.1$-$1$, and we fix the Sersic index $n$ to $1$, which is suitable for star-forming galaxies and also adopted by \citet{Yang2022b} and \citet{Morishita2018}. This latter choice is consistent with the median value that we find for a subset of sources with higher S/N for which the fit converges to a finite $n$ and $r_e$ when leaving all the parameters free (see also Mc Grath et al. in prep.).
The uncertainties on the sizes were estimated following \cite{Yang2022b} and re-scaled to the S/N from the photometry. 
The results obtained with \textsc{Galight} are robust, as shown by previous works \citep[e.g.,][]{Kawinwanichakij2021}, and in agreement with those estimated using traditional softwares such as \textsc{Galfit} \citep{Peng2002}.

\begin{figure}
\includegraphics[width=\linewidth]{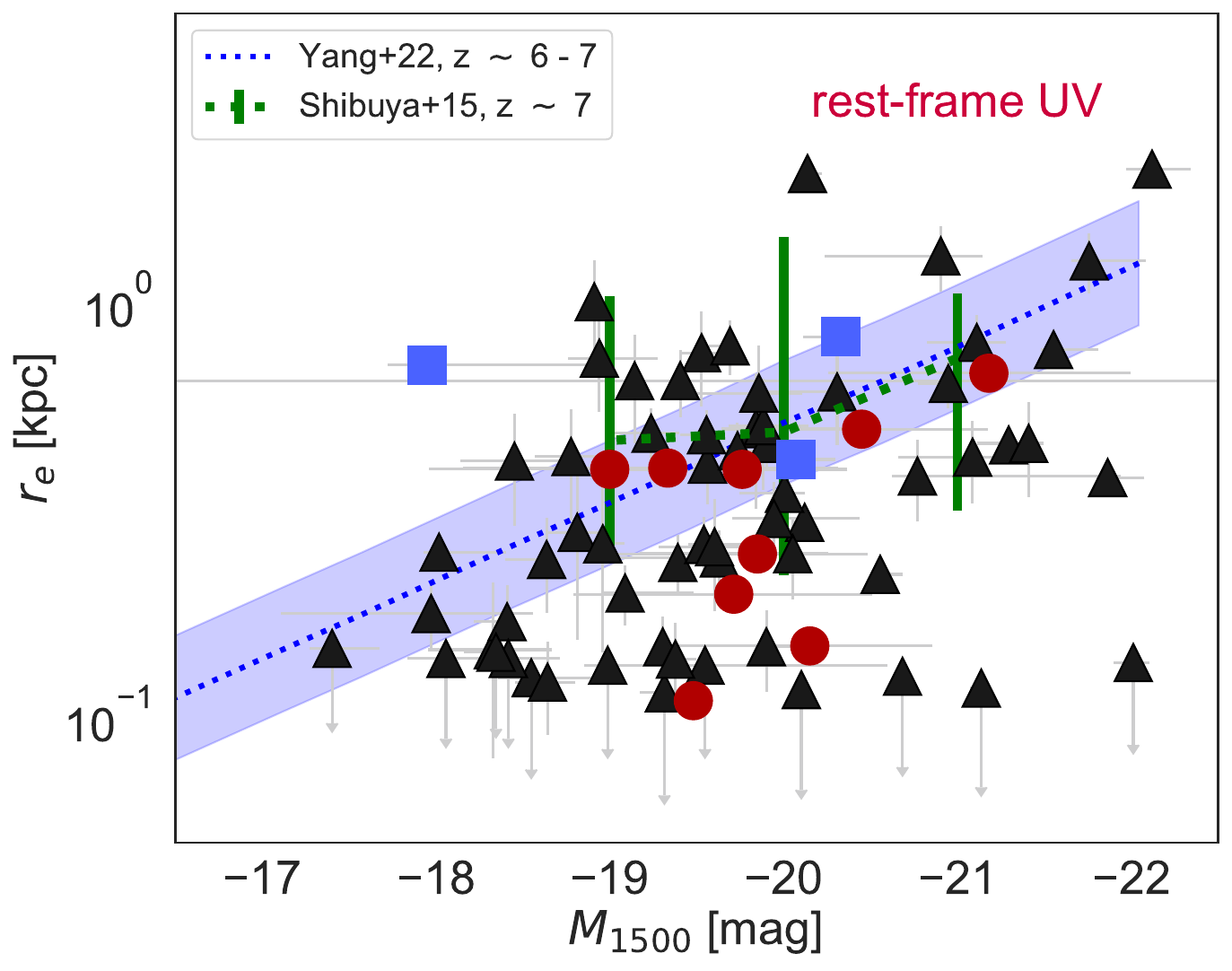}
\caption{Rest-frame UV $r_e$ vs. $M_{1500}$. Symbols are the same as in Fig. \ref{fig:beta_muv}. The green line shows the relation derived from HST data by \cite{Shibuya2015} at $z \sim 7$, the blue line shows the size–luminosity relation at $z \sim 6-7$ from \cite{Yang2022}. 
 \label{fig:re_muv}}
\end{figure}

For 50 galaxies, we used the \NIRCam\ photometry to measure $r_{e}$ in the F150W band (except for ID:542, for which the size is measured in F200W to improve the fit precision), corresponding to the UV rest-frame of the galaxies. For 19 galaxies where only the \hst\ photometry was available, we measured $r_e$ using the F160W filter, which has the highest S/N. 14 sources have profile resolutions that are likely unresolved, so we place an upper limit (see Calabrò et al. in prep). In cases where additional sources are present in the same cutout of a galaxy, we masked them or fitted them with additional Sersic profiles. We list all these measurements in in Table \ref{tab:summary1}.
To determine the minimum size measurable in the F150W band, we followed a similar approach to that recently adopted by \cite{Akins2023} in the F444W band. In brief, we performed a set of simulations by creating mock F150W images of galaxies (as observed by CEERS) with a Sersic profile, different magnitudes (from $25$ to $28$), and different intrinsic sizes from $0.005 \arcsec$ to $0.1 \arcsec$, in steps of $0.005$. We then applied PSF fitting with \textsc{Galfit}, considering unresolved a source if it is undetected (S/N $<2$) in the residual image. This procedure yields a minimum measurable size of $0.025 \arcsec$ (i.e., $\sim 123$ pc at redshift $8$), which we adopt in this work as a lower limit. We will describe these simulations in more detail in Calabro et al. 2023 (in prep.). 
As for the galaxies taken from previous works, for the M23 sample $r_{e}$ was measured in the F115W band; for the ERO sample, F200W was considered for the sources at $z>7$, and F150W for the galaxy ID:5144 at $z = 6.381$ \citep{Trussler2022}.
Typical sizes of our galaxies range from 0.1 to 2 kpc and are consistent with rest-frame UV $r_e$ measured during reionization by recent works \citep[][]{Morishita2023, Yang2022b, Shibuya2015}. In Fig. \ref{fig:re_muv} we show the relation between our measured $r_e$ and $M_{1500}$. Apart for a few outliers, we recover the well known magnitude-size relation: although with a large scatter, our results are consistent with the relation found at $z\sim7$ derived from HST data by \cite{Shibuya2015}, and the relation at $z\sim 6-7$ from \cite{Yang2022} based on photometrically selected galaxies lensed by six foreground Hubble Frontier Fields (HFF) clusters.
We note that most potential cosmic reionizers should have very small UV rest-frame dimensions ($\leq$ 0.4 kpc), indicating highly concentrated star formation as for example found by \cite{Flury2022b} and in a few intermediate redshift leakers such as Ion1 \citep{Ji2020}.

\subsection{AGN contamination}
While we recognize that AGN may also play some role in reionization, e.g., \cite{Madau15, Smith2018, Smith2020}, a concern with our current dataset is that any AGN identified here may constitute too small a sample, and might be too heterogeneous to properly evaluate their role in reionization. Therefore we exclude them in the current work, in order for us to provide the most robust measurements of the contribution of galaxies (non-AGN) to reionization, while the role of AGN is deferred to future studies with more suitable samples.
We first visually examined all spectra to see if there were any broad lines in them. Then, we employed the optical rest-frame spectroscopic diagnostics to distinguish between star-forming galaxies and AGNs. Most of our sources from the CEERS program have redshifts higher than 6.7 (7.07), so their \ha\ and \nii\ emission lines cannot be identified due to the long-wavelength limit of \NIRSpec\ G395M at $z \geq 6.7$, and $z \geq 7.07$ for the PRISM.  In any case,  at lower redshift \ha\ + \nii\ cannot be resolved with the PRISM. 
For this reason, we employed the mass-excitation (MEx) diagram \citep{Juneau2011, Juneau2014} with the division line identified by \cite{Coil2015} for $z=2.3$ galaxies and AGN from the MOSDEF survey, as already done in M23. According to the visual inspection and the position of our sources in the MEx diagram, we conclude that our sample contains one AGN (ID: 1019) and 69 star-forming galaxies. The AGN at $z=8.679$ was already identified and discussed by \cite{Larson2023}. 

 \section{Results}\label{sec:analysis}

 \subsection{Evaluating $f_{esc}$}

Assuming that the mechanisms that drive the escape of LyC photons are the same at all redshifts and depend only on the physical properties of the sources, several authors have recently attempted to derive empirical relations between $f_{esc}$ values and other observable and/or physical properties that can be measured also at high redshift. In particular, \cite{Lin2023} have applied the relation with the $\beta$ slope derived by \cite{chisholm2022}, while \cite{Saxena2023b} applied the relation predicted by \cite{Choustikov2023}, which relies on the $\beta$ slope, the $E(B - V)$, the \hb\ line luminosity, the $M_{1500}$, the R23, and the O32.
\\
In M23 we presented our own empirical relation calibrated on the \cite{Flury2022} low-redshift Lyman Continuum survey (LzLCS) sample, between $f_{esc}$ and $\beta$ slope, $r_e$, and $\log$O32 (Eq. 1 in M23). Due to the fact that O32 and $EW_0(\hb)$ exhibit a very tight correlation (Spearman correlation between them > 0.9), in M23 we used only one of the two values. However, since in some cases \hb\ is measurable while O32 is not, here we also present an alternative relation using $r_e$, $\beta$, and the $EW_0(\hb)$. This relation can be used when it is not possible to derive O32 due to a lack of one of the two lines. This new relation has the following form:
\begin{equation}
\log_{10}(f_{esc}) =A+ B \mathrm{EW}_0(\hb)+ C r_e + D \beta,
\label{eq}
\end{equation}
with $A = -1.92 [-2.46, -1.75]$, $B = 0.0026 [0.0019, 0.0035]$, $C = -0.94 [-1.14, -0.67]$, $D = -0.42 [-0.59, -0.33]$, where the values between the parentheses are in the 95th percentile distribution. In Appendix 2 we  present an analysis of the residuals between the measured $f_{esc}$ values  for the LzLCS sample and those predicted using both relations.

\begin{figure*}
\centering
\includegraphics[width=0.48\linewidth]{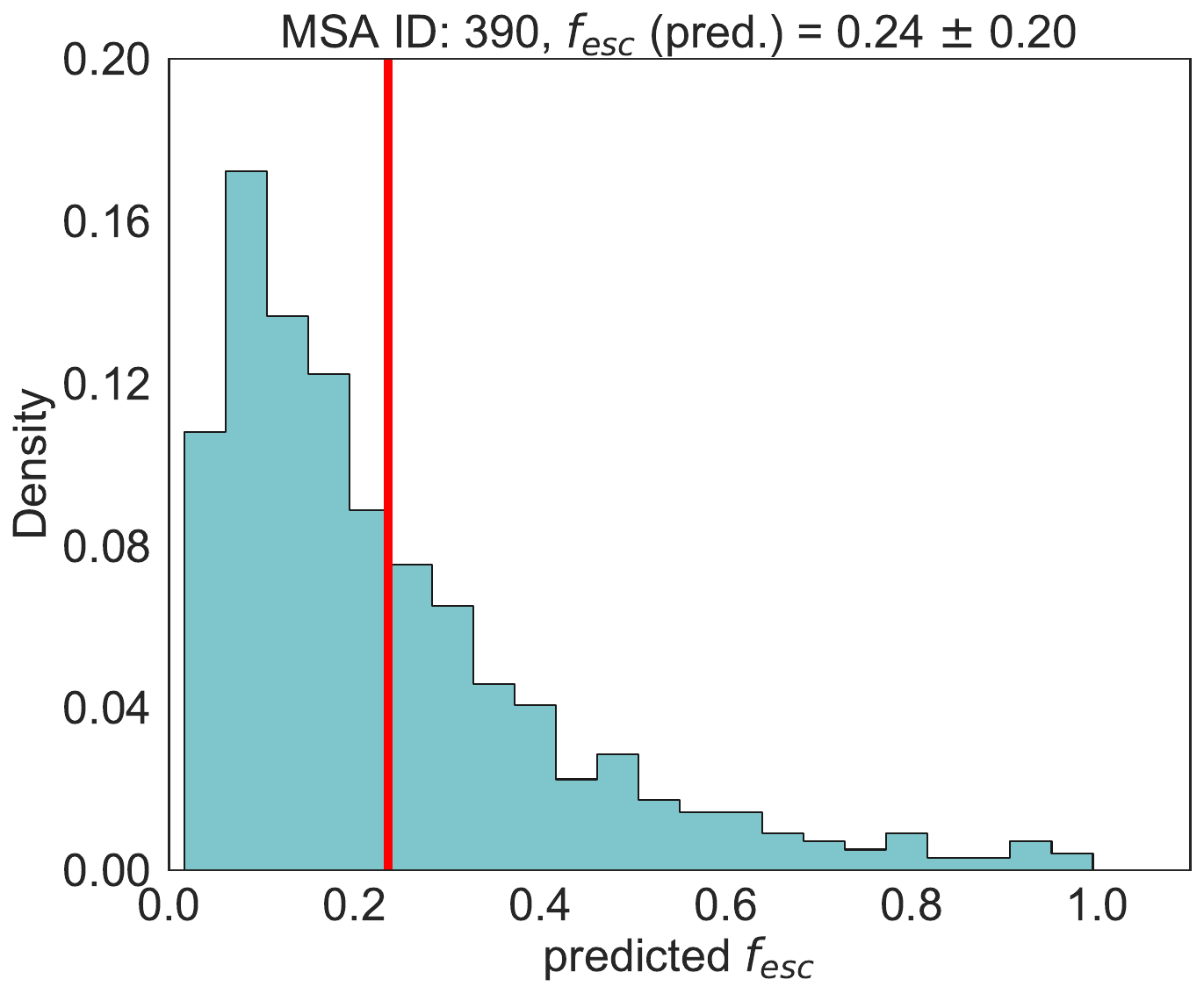}
\includegraphics[width=0.5\linewidth]{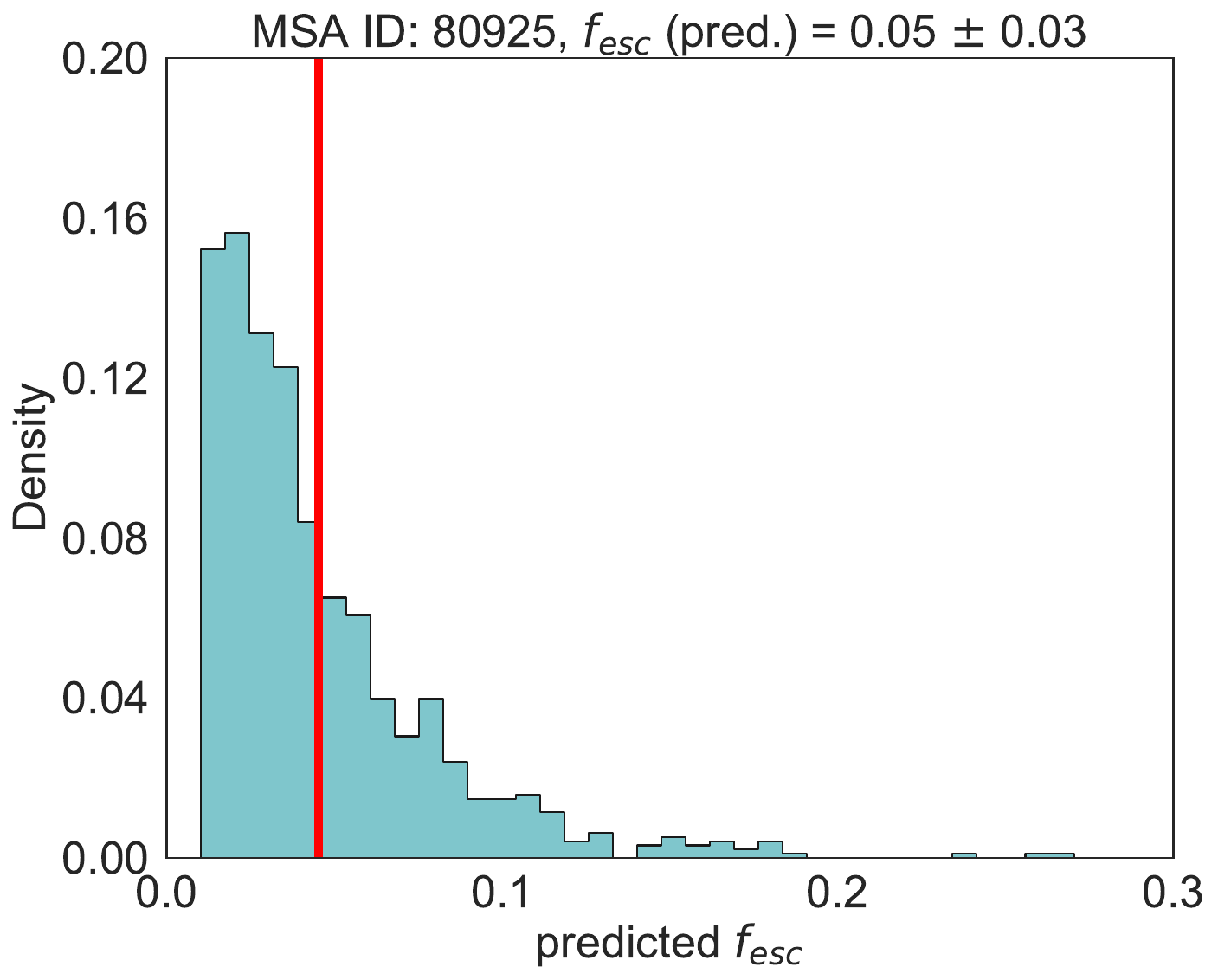}
\caption{PDFs of the $f_{esc}$ values for two sources from the CEERS sample (MSA IDs: 390 and 80925). The considered $f_{esc}$ of the source, which is the mean $f_{esc}$ of the distribution, is shown in red.
 \label{fig:fesc_example}}
\end{figure*}

Using either Eq. 1 in M23 or Eq. \eqref{eq}, we predicted the $f_{esc}$ value for the CEERS 65 star-forming galaxies, in addition to the  GLASS+DDT and  ERO sources for which we have the $\beta$ slopes. As already mentioned, the UV half-light radius of 1 source from the CEERS sample could not be determined due to the inability to achieve a good fit of the profile. Moreover, in 2 cases, the $\beta$ slope could not be measured. Since these quantities appear in both of the proposed equations, we were unable to estimate $f_{esc}$ for 3 sources of the CEERS sample. For the remaining sources, we used the M23 equation in 49 cases in which O32 is measured accurately or it is a limit but \hb\ is not evaluated, and the Eq. \eqref{eq} for the other 13 cases. In total, we predict an $f_{esc}$ value for 74 sources from the three samples. Given the uncertainty both on the coefficients of the relations and on the quantities on which $f_{esc}$ depends, we estimate the $f_{esc}$ errors using a bootstrap method. We use $n = 1000$ Monte Carlo simulations varying both the coefficients and the individual properties within their uncertainty. The results provide an $f_{esc}$ distribution from which we determine the mean $f_{esc}$ and the standard deviation for each galaxy, which is taken as the uncertainty. In Fig. \ref{fig:fesc_example} we show two examples of the probability distribution function (PDF) of the $f_{esc}$ values resulting from the above Monte Carlo runs, for a galaxy with modest inferred mean $f_{esc}$ (0.05) and one with a high inferred  mean $f_{esc}$ (0.24). In  Table \ref{tab:summary1} we report the mean $f_{esc}$ and the standard deviation for the CEERS galaxies, in Appendix 1, Table \ref{tab:summary} we report the same values for the GLASS and ERO sources.

\begin{figure}
\includegraphics[width=\linewidth]{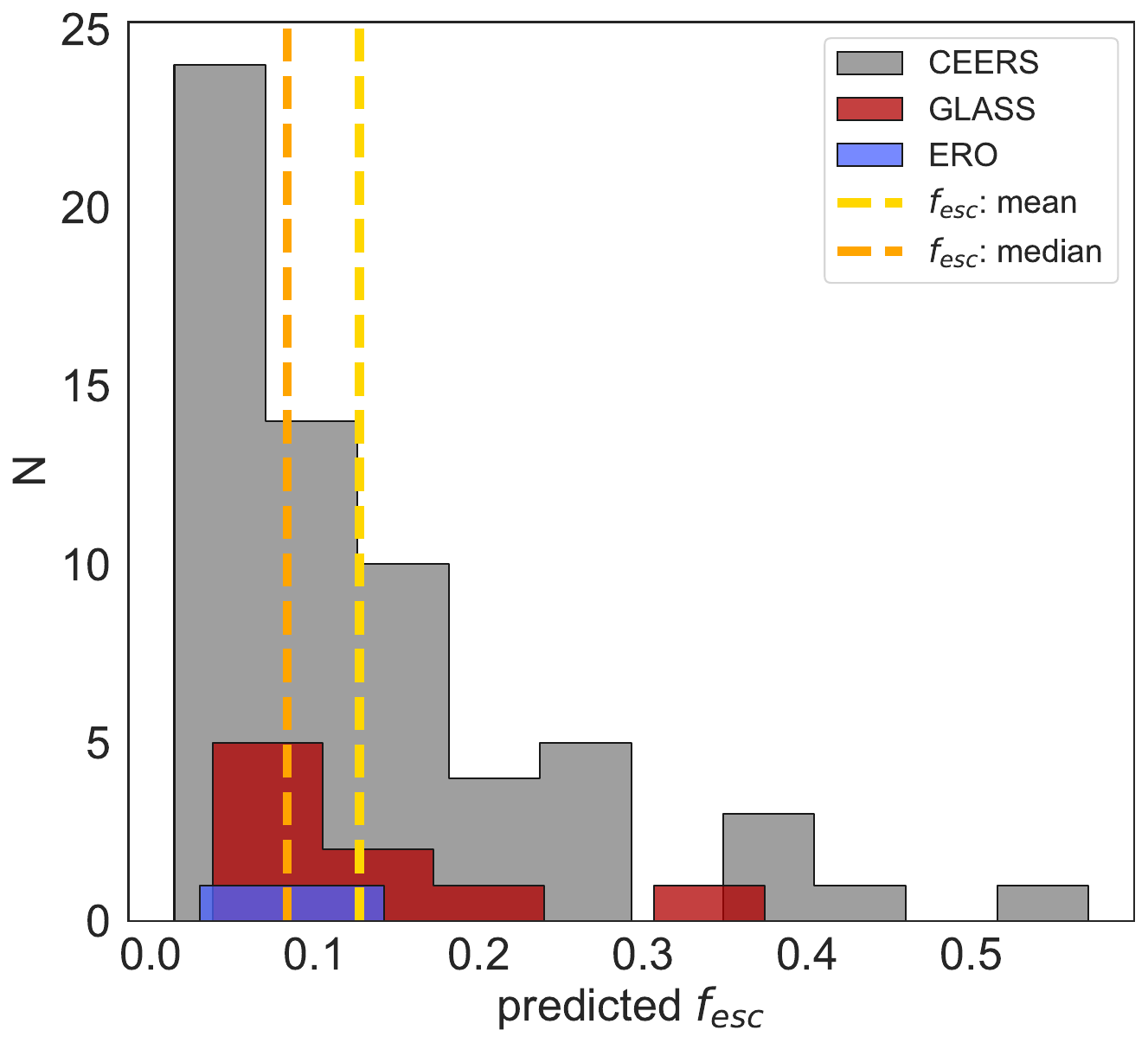}
\caption{Predicted $f_{esc}$ distribution for the analyzed sources at $6 \leq z \leq9$ (grey: CEERS sample; red and blue are respectively the GLASS sample and the ERO sample). The mean $f_{esc}$ of the sample is shown in yellow, the median $f_{esc}$ is presented in orange.
 \label{fig:fesc_distribution}}
\end{figure}

In Fig. \ref{fig:fesc_distribution} we present the distribution of the inferred mean $f_{esc}$ values. Most of our galaxies have modest inferred $f_{esc}$, of the order of 0.10 or below.  The average $f_{esc}$ for our sample (with the standard error of the mean) is $0.13 \pm 0.02$. This value is affected by the high $f_{esc}$ ($0.3-0.5$) inferred for a handful of sources. The median in this case is a more representative value and it is equal to $0.08 \pm 0.02$. 
To evaluate the impact of using the mean $f_{esc}$ for each galaxy instead of the full PDF (which is not gaussian but more lognormal), we produced the same distribution shown in Fig. \ref{fig:fesc_distribution}, this time  stacking  the individual PDF of all galaxies. The resulting distribution essentially unchanged: computing the mean and median values they are respectively 0.11 and 0.08, confirming that our results are robust.

\subsection{$f_{esc}$ dependencies}

\begin{figure*}
\includegraphics[width=0.487\textwidth]{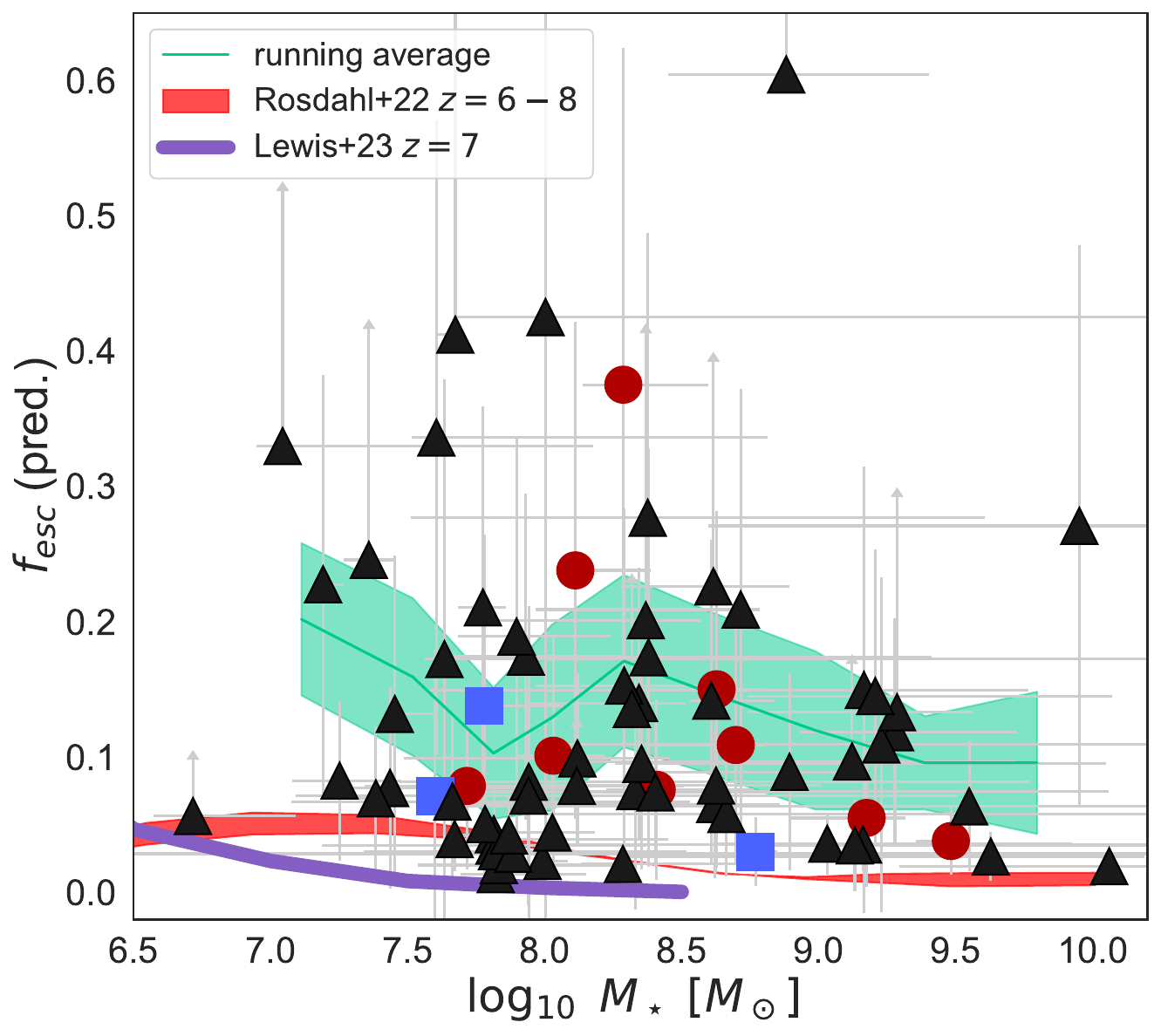}
\includegraphics[width=0.5\textwidth]{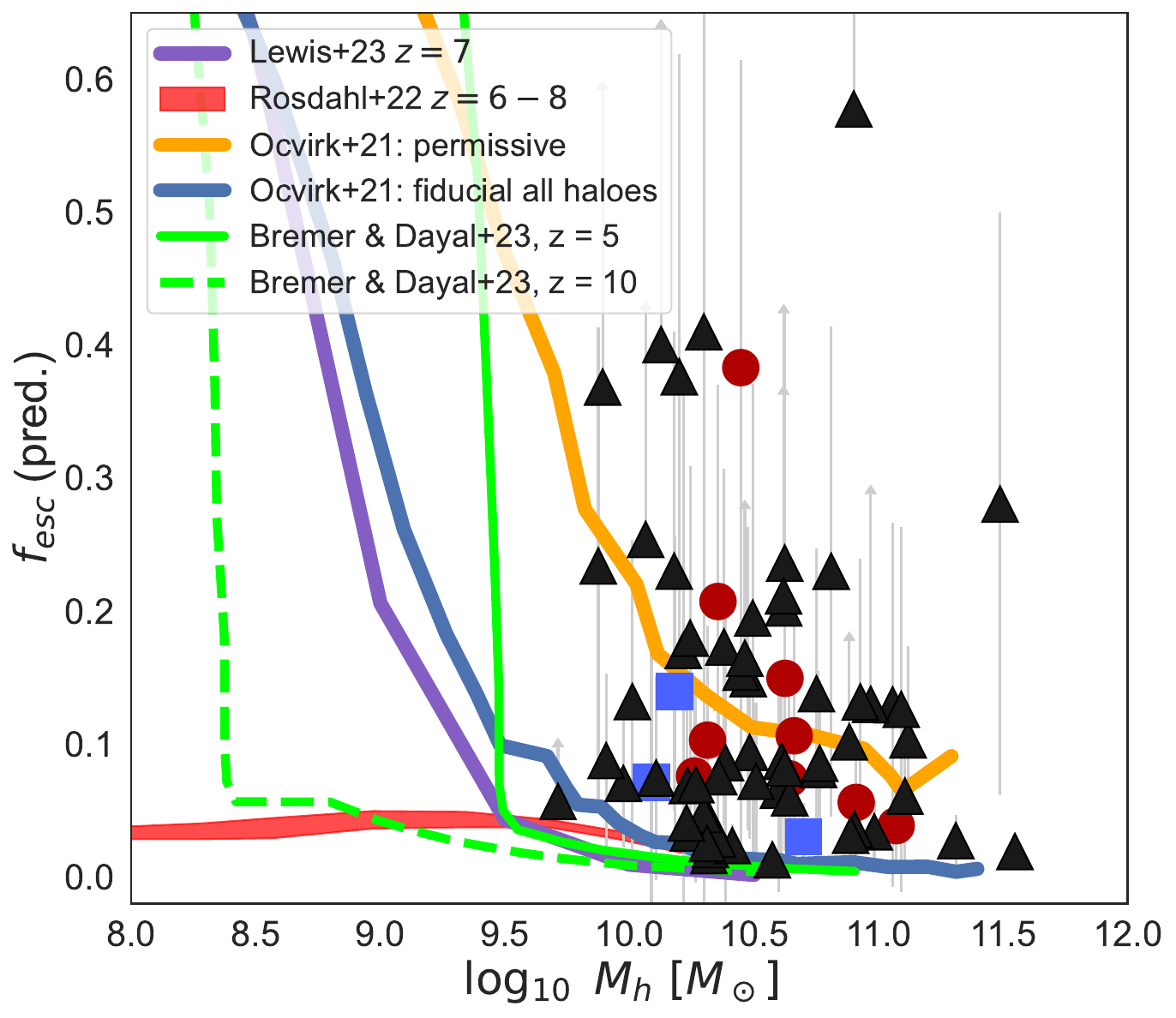}
\caption{Left: predicted $f_{esc}$ vs  stellar mass ($\log_{10} M_\star$). Symbols are the same as in Fig. \ref{fig:beta_muv}. The green line shows the running average for our sample, while the red one is the prediction at $z = 6-8$ from \cite{Rosdahl2022}. Right: predicted $f_{esc}$ vs halo mass ($\log_{10 }M_h$) estimated with \cite{Behroozi2019} conversion. The blue and orange lines are the models from \cite{Ocvirk2021}, the violet one is the prediction from \cite{Lewis2023}, the line ones are the predictions from \cite{Bremer2023}.
\label{fig:fesc_mass}}
\end{figure*}

In Fig. \ref{fig:fesc_mass}, left panel, we plot the predicted $f_{esc}$ values versus the stellar mass $M_*$. We show average binned values (using a running average) with the shaded area indicating the $1\sigma$ uncertainty. We find that low-mass galaxies tend to have slightly higher escape fractions, although the relation is rather scattered. For comparison we also plot the prediction by \cite{Rosdahl2022} based on SPHINX cosmological radiation-hydrodynamical simulations of reionization. Their simulated values of $f_{esc}$ are generally lower than our predictions, well below 0.1 during most of the EoR, although they also find the same dependence on total stellar mass.

Since most simulations predict $f_{esc}$ versus halo mass $M_h$ relations, we converted our stellar masses $M_*$ into halo masses $M_{h}$ following the relation as a function of redshift derived by \cite{Behroozi2019}. We plot the $M_{h}$ versus the predicted $f_{esc}$ values in 
Fig. \ref{fig:fesc_mass}, right panel. 
We compare our results to the prediction by \cite{Rosdahl2022} (see above) and to those obtained by \cite{Ocvirk2021} and  \cite{Lewis2023} using  RAMSES-CUDATON radiation-hydrodynamical simulations. 
These simulations aim at reproducing the observed \lya\ opacity distribution. Their predicted $f_{esc}$ are $\sim 1$ for very low-mass galaxies and drop at $M_h \geq \times10^{9.5} M_\odot$. We plot both the fiducial and "permissive" model of \cite{Ocvirk2021}, where this second one allows a more permissive recipe for SF also above the temperature of $T_*=2\times 10^4 K$. In \cite{Lewis2023} the fiducial model of \cite{Ocvirk2021} is extended through the inclusion of a physical model for dust production, coupled to the radiative transfer module. 
Finally we plot the predictions by \cite{Bremer2023} that are based on DELPHI simulations at $z =5$ and $z = 10$. In this work, reionization starts at $z\sim 16$, is complete at $z = 5.67$ and it is dominated by faint, low-mass galaxies with $M_{h} \leq 10^{7.8} M_\odot$ at $z \sim 15$ that show $f_{esc}$ up to 0.7.
\\
Most of the above models predict a very rapid increase of $f_{esc}$ with  decreasing halo mass, below $M_h \simeq 10^{10} M_\odot$, a range which we barely sample with our observations, and a very low almost null $f_{esc}$ for the more massive halos, at odds with our inferences. In the range of halo mass observed, simulations are more than 1 $\sigma$ away from our inferred $f_{esc}$.
\\
The strong discrepancy between the $f_{esc}$ values we derive from \NIRSpec\ data and the model predictions could be due to a number of aspects:
\begin{itemize}
    \item{It may be that simulations do not adequately capture the bursty nature of star formation. It has been shown that Supernova (SN) feedback plays a critical role in creating regions with higher transparency for LyC escape.  As a consequence in the models there is  a positive correlation between $f_{esc}$ and the SFR measured over the last 10 million years \citep{Rosdahl2022}. This suggests that bursty star formation contributes to higher $f_{esc}$ values. However accurately quantifying the burstiness of star formation observationally, and comparing it to a simulation's burstiness is a difficult task. 
    For instance, it has been suggested that H$\alpha$ / FUV fluxes for could help quantifying SFR burstiness observationally \citep{sparre2017_burstiness}, but this requires a fairly sophisticated post-treatment of simulations, and is very sensitive to the details of star formation, feedback (SN and radiative) and ISM modelling.  Other probes of burstiness have been and will be proposed \citep{sun2023_burstiness}, and may offer avenues of progress on this topic.}
    
\item{Another potential reason for the large discrepancy could be the description of the thermodynamical state of the shock-heated multi-phase CGM \citep{vandeVoort2015}. For example, in a case in which a clumpy CGM is composed of 
hot, highly ionized  gas surrounding cold dense clumps,  if the cold phase is sufficiently dense and the hot phase has high pressure, the clumps may have a small cross-section: with a small total covering fraction, a high $f_{esc}$ value could be observed. Insufficient spatial resolution in this case would imply artificially larger clumps, leading to a higher covering fraction and reducing the $f_{esc}$. The complexity of this behavior is being explored in simulations \citep{Gronke2020}.}
\item{Finally we should keep in mind that the relations that we have used to infer the $f_{esc}$ for galaxies in the EoR have been derived and tested  using the LzLCs sample that is located at low redshift ($z=0.2-0.4$). Therefore its applicability to the CEERS sample ($6 \geq z \geq 9$}) is not straightforward.  The large discrepancy with simulations might be due to an overestimate of the $f_{esc}$ in the EoR. 
\end{itemize}

The permissive model of \cite{Ocvirk2021} is the only one that has average $f_{esc}$ high enough to be comparable to our values. This can be attributed to the permissive run's unique characteristic of permitting star formation in cells with temperatures potentially exceeding $2 \times 10^4$ K. These higher temperatures inherently lead to greater ionization and increased transparency compared to the fiducial run and hence to larger values of $f_{esc}$. Interestingly, this model is {\em not} the one favored by \cite{Ocvirk2021} as it leads to an overionization of the Lyman-$\alpha$ forest characterized by unrealistically low Lyman-$\alpha$ IGM opacities.

\begin{figure}
\includegraphics[width=\linewidth]{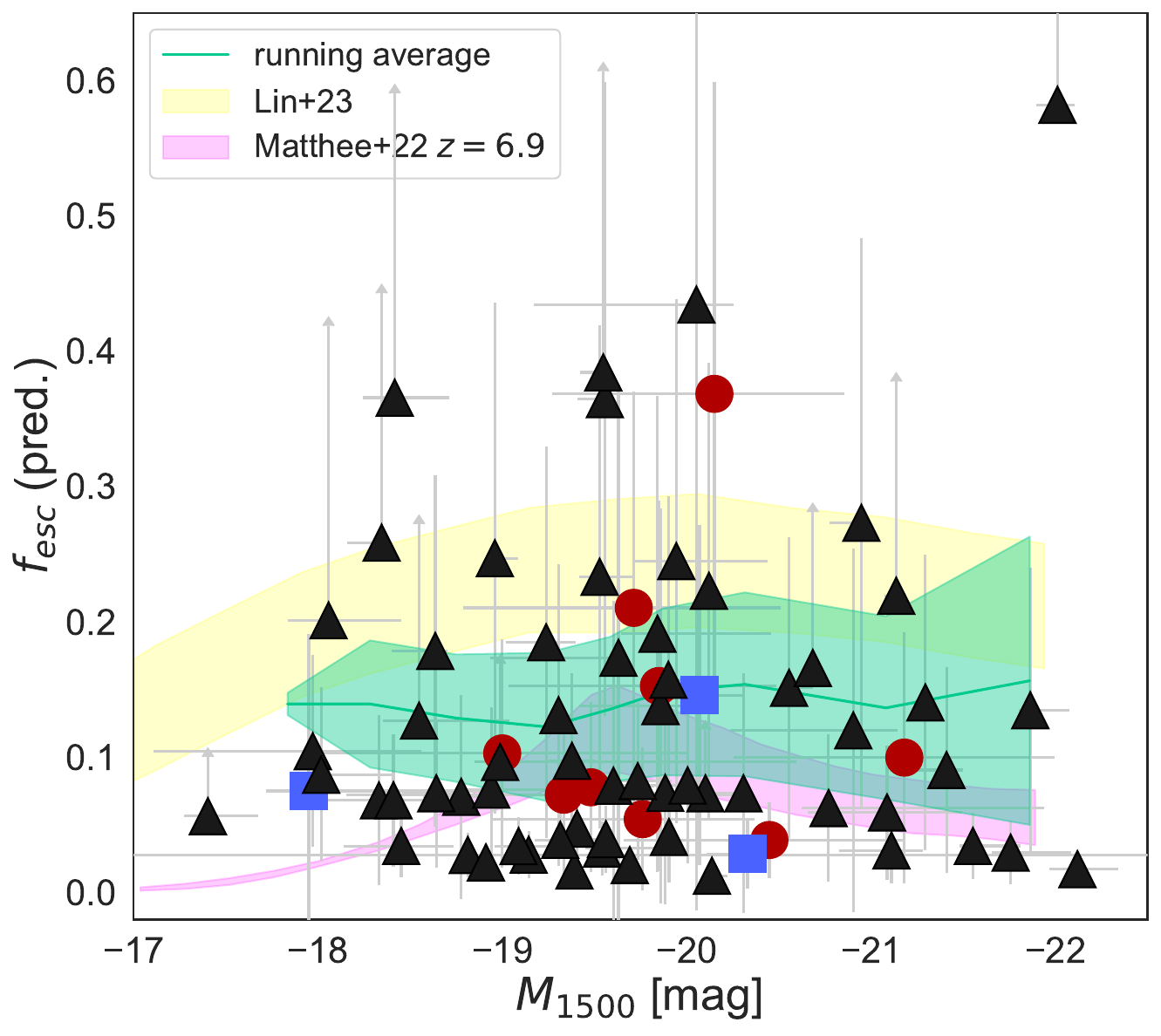}
\caption{Predicted $f_{esc}$ vs $M_{1500}$. Symbols are the same as in Fig. \ref{fig:beta_muv}. The green line shows the running average for our sample, the pink one refers to \cite{Matthee2022} and the yellow one to \cite{Lin2023}. 
\label{fig:fesc_muv}}
\end{figure}

In Fig. \ref{fig:fesc_muv} we plot our predicted average $f_{esc}$ values versus the UV magnitude $M_{1500}$. We note that Eq. \eqref{eq} and the M23 relation have been derived on the LzLCS which only contains galaxies brighter than $M_{1500} \simeq -18.5$. Therefore for our few faintest objects using the above equations might be an incorrect extrapolation.
Our average $f_{esc}$ is almost constant within the observed magnitude range, altough we point out that 
we might start to be biased at the faintest luminosities  (especially for objects with faint emission lines and hence small $f_{esc}$) due to the spectroscopic flux limit of the CEERS survey. 
In the same Figure we also show the predicted $f_{esc}$ vs $M_{1500}$ relationship by \cite{Lin2023}, who analyzed 3 galaxies at $z\geq 8$ behind the cluster RX J2129.4+0009. They developed an empirical model based on the LzLCS program, which first defines for a given galaxy a probability of being a LyC-leaker based on $M_{1500}$, O32 and $\beta$ and then infers the $f_{esc}$ values from the $\beta$ slope following \cite{chisholm2022}. They predict for bright galaxies a very flat relation, similar to ours but with $f_{esc}$ values that are about a factor of 2 larger. In addition, they predict that $f_{esc}$ should slowly decrease for galaxies fainter than $M_{1500}=-19$. Essentially this is due to the fact that the probability of faint galaxies of being LCE becomes lower. However they extrapolate this result from the LzLCS, which as already mentioned earlier, contains no sources below $M_{1500} = -18.5$. 
As a final comparison, we plot the results by \cite{Matthee2022} who produced a semi-empirical model based on constraints on the escape fractions of bright LAEs at $z \sim 2$. These authors find that $f_{esc}$ peaks between $-19 \leq M_{1500}\leq -20$ and then decreases very rapidly at fainter magnitudes (the so-called reionization by the oligarchs). At magnitudes brighter than $M_{1500}=19$ our average results are consistent with theirs, within the uncertainties, but we do not observe the strong decrease at fainter  magnitudes.

\subsection{Redshift evolution}
\begin{figure*}
\includegraphics[width=\linewidth]{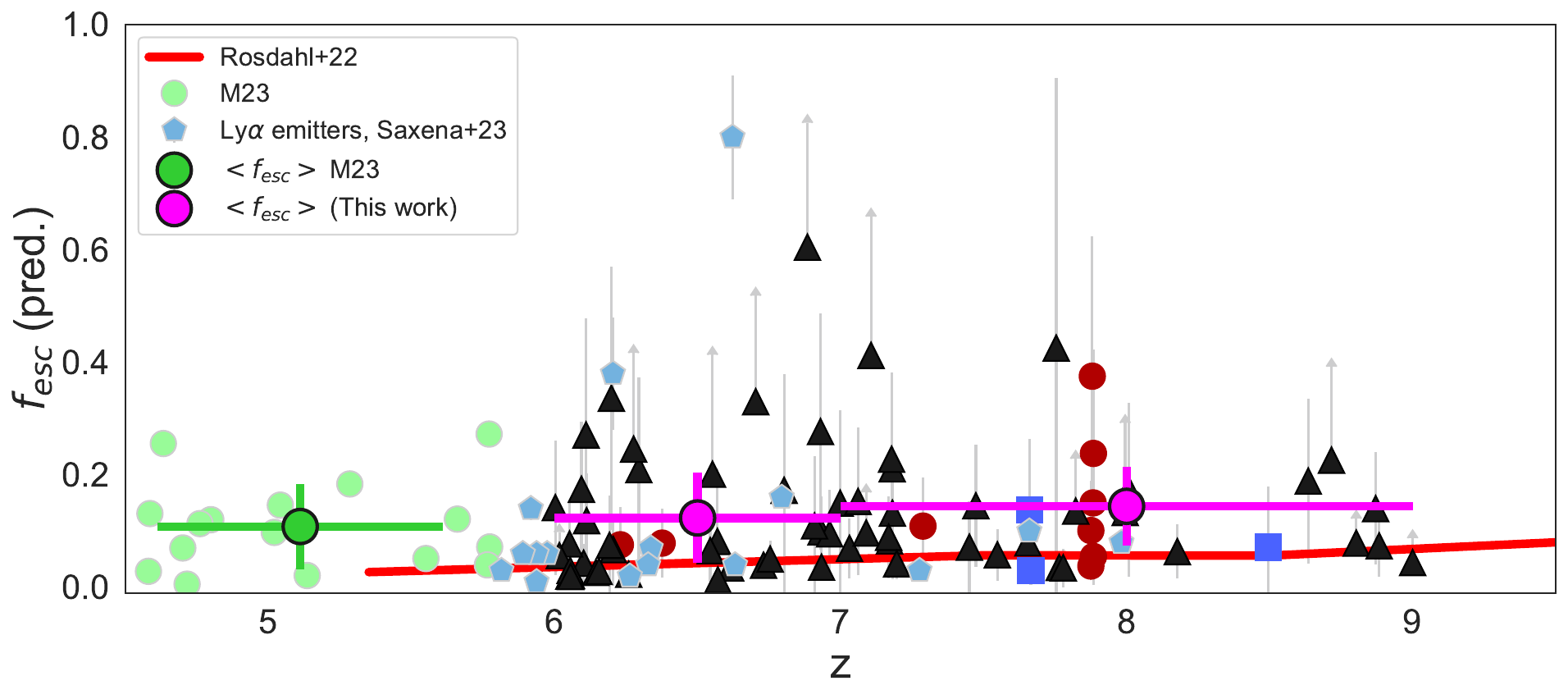}
\caption{Predicted $f_{esc}$ as a function of redshift. Symbols are the same as in Fig. \ref{fig:beta_muv}. The magenta points are the average $f_{esc}$ values of our sample at $z = 6.5$ and $z = 8$. The green dots are the M23 sources, while the green point is the average $f_{esc}$ value at $z = 5.1$. The blue hexagons represent the \lya\ emitters from \cite{Saxena2023b}. The red line is the prediction at fixed $M_{1500} = - 19$ from \cite{Rosdahl2022}. 
\label{fig:fesc_z}}
\end{figure*}

In Fig. \ref{fig:fesc_z} we plot our predicted $f_{esc}$ versus the redshift. We also plot the sources from M23 at redshift lower than 6 which were derived with the same method. The average $f_{esc}$ in the three redshift bins, $5 \leq z < 6$, $6 \leq z < 7$ and $7 \leq z \leq 9$, are respectively equal to 0.11, 0.12 and 0.14. We therefore observe a slight increase of the average $f_{esc}$ with redshift, although statistically not significant. A similar trend would be observed using the median values. We also show the predicted $f_{esc}$ as function of redshift from \cite{Rosdahl2022}, derived from Figure 6 of their paper at a median $M_{1500} = -19$. As previously discussed, their $f_{esc}$ values are generally lower than ours, but they predict a slow increase of the $f_{esc}$ with redshift which is very similar to what we observe.

In the same plot we also show the sample of \lya\ emitters at $z=6-8$ from the JWST Advanced Deep Extragalactic Survey (JADES) presented in \cite{Saxena2023b}, which span the same $M_{1500}$ range as our sources. They predict the $f_{esc}$ using an equation proposed by the \cite{Choustikov2023}, based on the SPHINX simulations which uses six observed galaxies' properties to infer the angle-averaged (and not sight-line dependent) $f_{esc}$. 
We see that non-\lya\ emitters and \lya\ emitters at the same redshift and in the same UV magnitude range do not show significant differences in the predicted $f_{esc}$, although determined using two different and independent methods. This might be a first indication  that in the EoR, when the visibility of \lya\ emission is increasingly suppressed by neutral IGM, the \lya\ line emerging from the galaxies is not a good indicator of the LyC photons' escape and therefore other indirect indicators are needed. Further investigation of this important issue is in progress and will be presented in  follow up paper.

\subsection{Extreme LyC emitters}
We analyzed in more details the 16 sources from our final sample that show an $f_{esc}$  higher than 0.2. The majority of them show an intense O32 or high EW(\hb) coupled with small $r_e$ or very blue $\beta$ slope. We highlight the fact that an extremely blue $\beta$ is a very good predictor of a high $f_{esc}$: 10 out of 17 sources with $\beta> -2.5$ have a predicted $f_{esc}> 0.2$. Indeed \cite{chisholm2022} identified the $\beta$ slope as one of the best indirect indicators. However this condition 
does not seem necessary, since there are several sources that have  more average  $\beta$ slopes (i.e., of the order of $-2$) but for which we predict high $f_{esc}$ because they are both extremely compact and have a high O32 or high EW(\hb). 
\textrm{Similarly of the 8 unresolved sources for which we are able to infer $f_{esc}$, 5 are extreme leakers ($f_{esc} >0.2$). However we have some leakers with $r_e$ larger than 0.5 kpc. }
Overall, there is not one single property that stands out as more important. This reinforces the idea that more than one indicator is needed to correctly identify the entire population of LyC emitters.

\subsection{Ionizing photon production efficiency}\label{sec:xi}

\begin{figure}
\includegraphics[width=\linewidth]{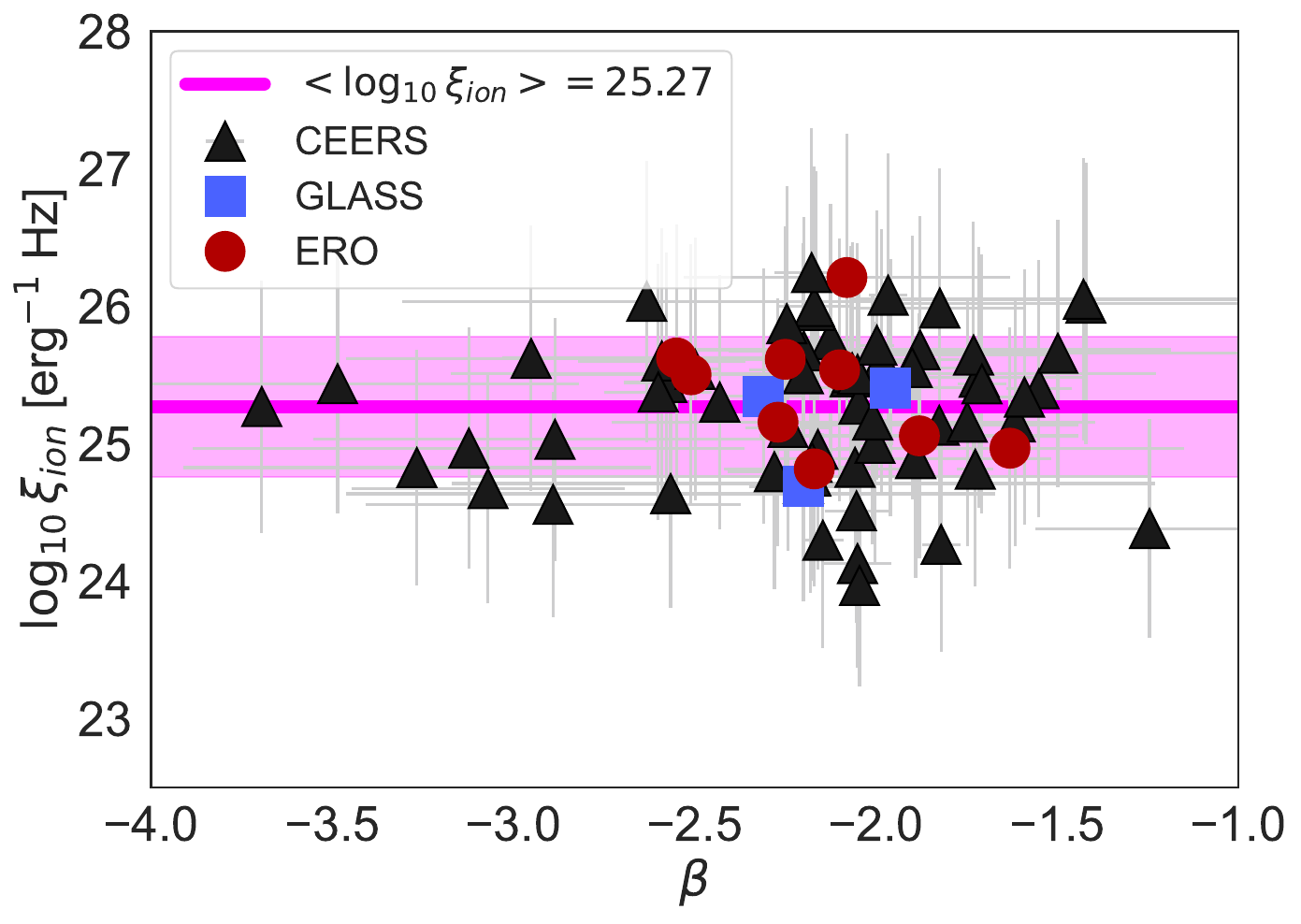}
\caption{$\log \xi_{ion}$ vs $\beta$. The magenta line shows the mean $\log \xi_{ion}$ for our sample.
 \label{fig:xi_ion}}
\end{figure}

Direct constraints on $\xi_{ion}$ can be obtained from the measurement of Balmer emission lines luminosity after correcting for dust attenuation \citep[e.g.][]{Schaerer2016,Shivaei2018} or from modelling the contribution of these optical emission lines to the broad band measurements when spectroscopic observations are not available \citep[e.g.][]{Bouwens2016}. Unfortunately, the \ha\ line is outside the observed range of most of our galaxies (see Sec. \ref{sec:data_line}) and the \hb\ line is also missing from some sources: in addition, there are still some calibration uncertainties on NIRSpec absolute flux (and therefore luminosity). \cite{Chevallard2013} showed that $\log\xi_{ion}$ can be measured by using EW(\oiiidoub) \citep[see also][]{Tang_2019}. The \oiii\ lines are clearly detected for all sources, and in addition the EW measurements have less calibration uncertainty compared to the flux.
We calculated the $\log\xi_{ion}$ values from $EW(\oiii)$ following the Eq. 3 from \cite{Chevallard2018}. 
We obtain an average $\langle \log\xi_{ion} \rangle = 25.27 \pm 0.51$, which is consistent with predictions from physical models \citep[][]{Yung2020b, Wilkins2016} and slightly lower than other measurements at the EoR. For example, \cite{Saxena2023b, Simmonds2023} estimated $\log \xi_{ion}$ from \ha\ luminosity, finding respectively average values of $25.56$ and $25.44$ although their samples included \lya\ emitters whose photon production efficiency is generally higher, while \cite{Castellano2022, Prieto-Lyon2022, Endsley2023}, using SED fitting, obtain an average value of $\log \xi_{ion}$ of 25.14, 25.33, 25.7 respectively. 
In Fig. \ref{fig:xi_ion} we show the distribution of $\xi_{ion}$ for our sample. We do not find any correlation with the $\beta$ slope: our best fit is consistent with the average value also shown in the Figure. At variance with this, \cite{Prieto-Lyon2022} find a slight dependence on this property for galaxies at $z=3-7$, in the sense that bluer star-forming sources tend to have higher photon production efficiencies \citep[see also][]{Castellano2023}. We also do not find any dependence of $\xi_{ion}$ on $M_{1500}$ in accordance to what found by \cite{Prieto-Lyon2022, Endsley2023}. Note that the recent results by  \cite{atek2023} indicate a higher $\xi_{ion }$   for much fainter galaxies ($M_{UV} \geq -17$) during the EoR.

\section{The ionizing photon production of bright and faint sources }\label{sec:discussion}

Having derived predictions for $f_{esc}$ and $\xi_{ion}$ for our large sample of galaxies in the EoR, our goal is now to solve Eq. \eqref{eq:N_ion} and determine the relative contribution of galaxies as a function of $M_{1500}$, to establish which sources contributed most to the total ionizing photon production rate at these epochs. We consider: 1) $\rho_{UV}$ from the Luminosity Function (LF) of \cite{Bouwens2021} at our median redshift $\langle z \rangle \sim 7.2$. The best-fit $\alpha$ slope that characterizes the faint-end of the UV-LF is $-2.06 \pm 0.03$; 2) $f_{esc}$ as a function of $M_{1500}$ from the values derived in Fig. \ref{fig:fesc_muv} between  $M_{1500}$ -22 and  -18 (i.e. the range covered by our observations); we use a fixed value of 0.10 at fainter magnitudes, where we have only few sources, and a value of 0.05 at magnitudes brighter than -22, where we do not have any observed source in our sample; 3) $\langle \log\xi_{ion} \rangle = 25.27$, which does not vary with $M_{1500}$, as found in Sec. \ref{sec:xi}.
 We assume a low luminosity cut at $M_{1500}=-13$ and a high luminosity cut of $M_{1500}=-23$  \citep[as in][]{Robertson2015}.
 \\
To estimate the total $\dot{n}_{ion}$  we proceeded as follows: we first discretized the $M_{1500}$ range over [-13,-23] in bins of width 1 mag. For each of these intervals we calculated $\rho_{UV}$ in the considered magnitude bin and multiplied it by the appropriate $\xi_{ion}$ and $f_{esc}$. We then summed these values  to estimate the total $\dot{n}_{ion}$. The total integrated ionizing emissivity at $z=8$ and $z =6$ are respectively $\log{\dot{n}_{ion}} = 50.50 \pm 0.38$ and $50.75 \pm 0.35 \ \text{s}^{-1} \text{Mpc}^{-3}$, consistent with the canonical threshold needed to maintain the Universe ionized at $z = 7$ \citep[e.g.,][]{Madau1999, Gnedin2022} and in the range of previous determination \citep{Finkelstein2019,Bouwens2015, Robertson2015}.


\begin{figure*}[ht!]
\begin{minipage}{0.5\textwidth}
\centering
\includegraphics[width=\linewidth]{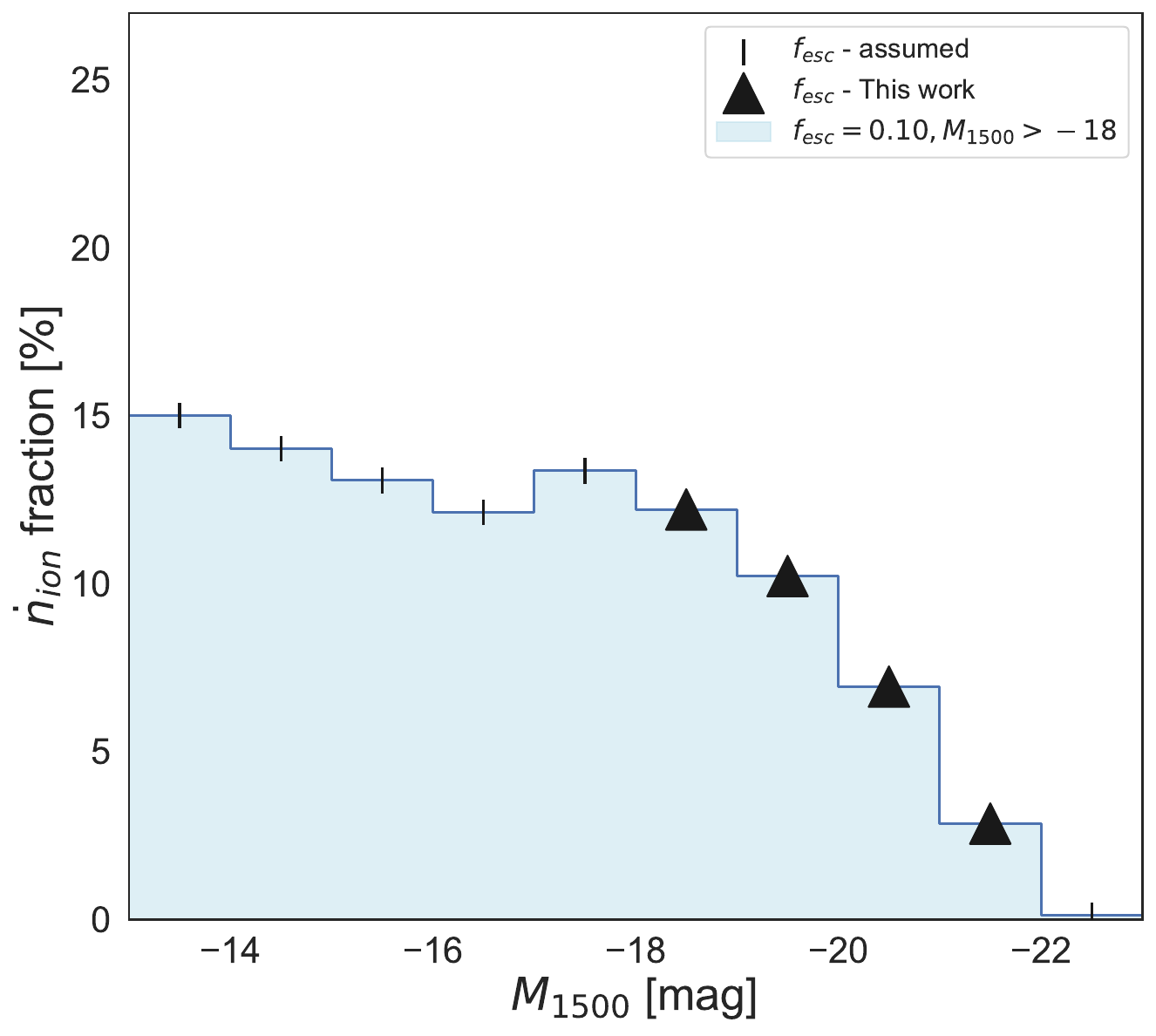}
\end{minipage}
\begin{minipage}{0.5\textwidth}
\centering
\includegraphics[width=0.48\linewidth]{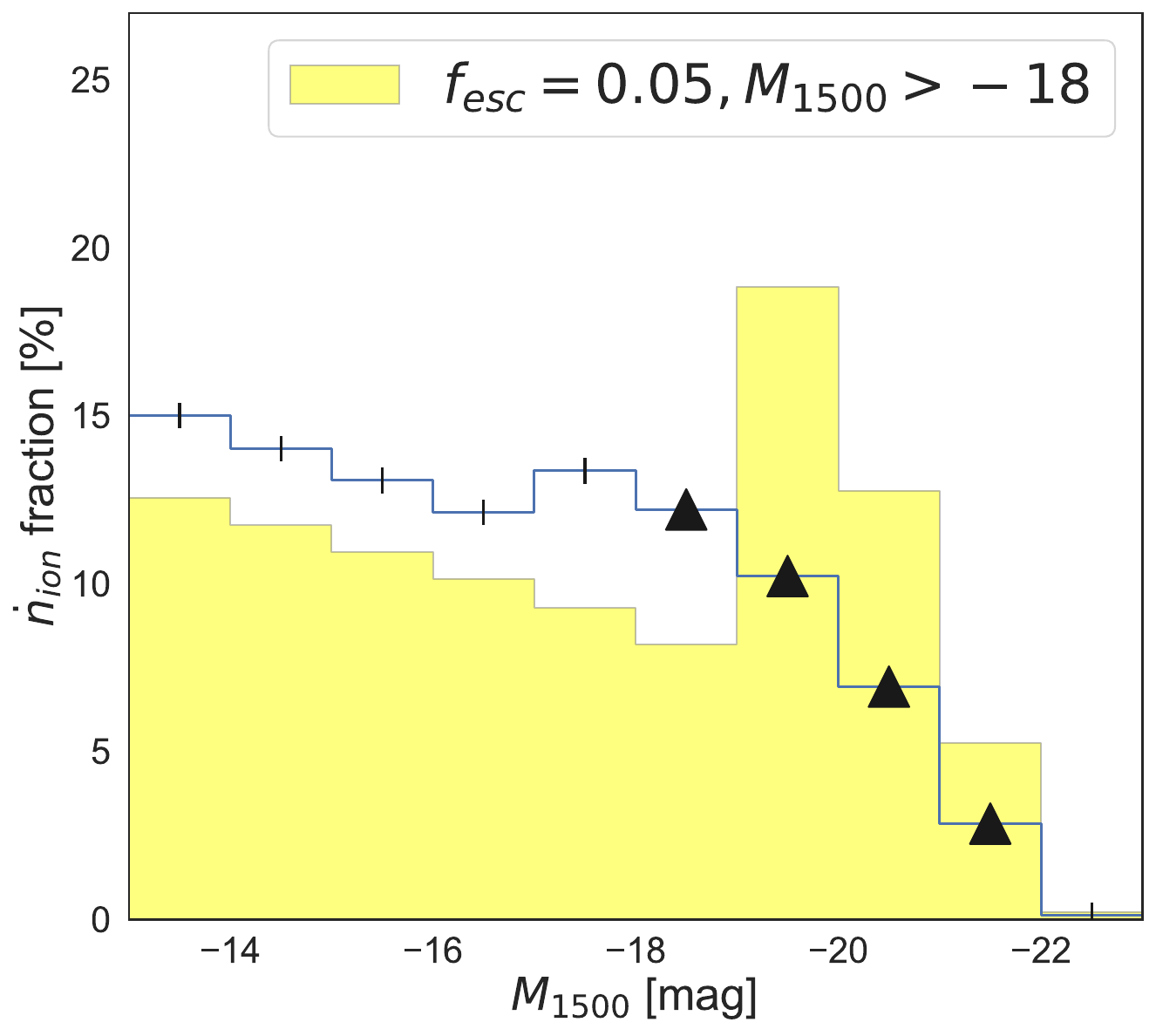}\quad
\includegraphics[width=0.48\linewidth]{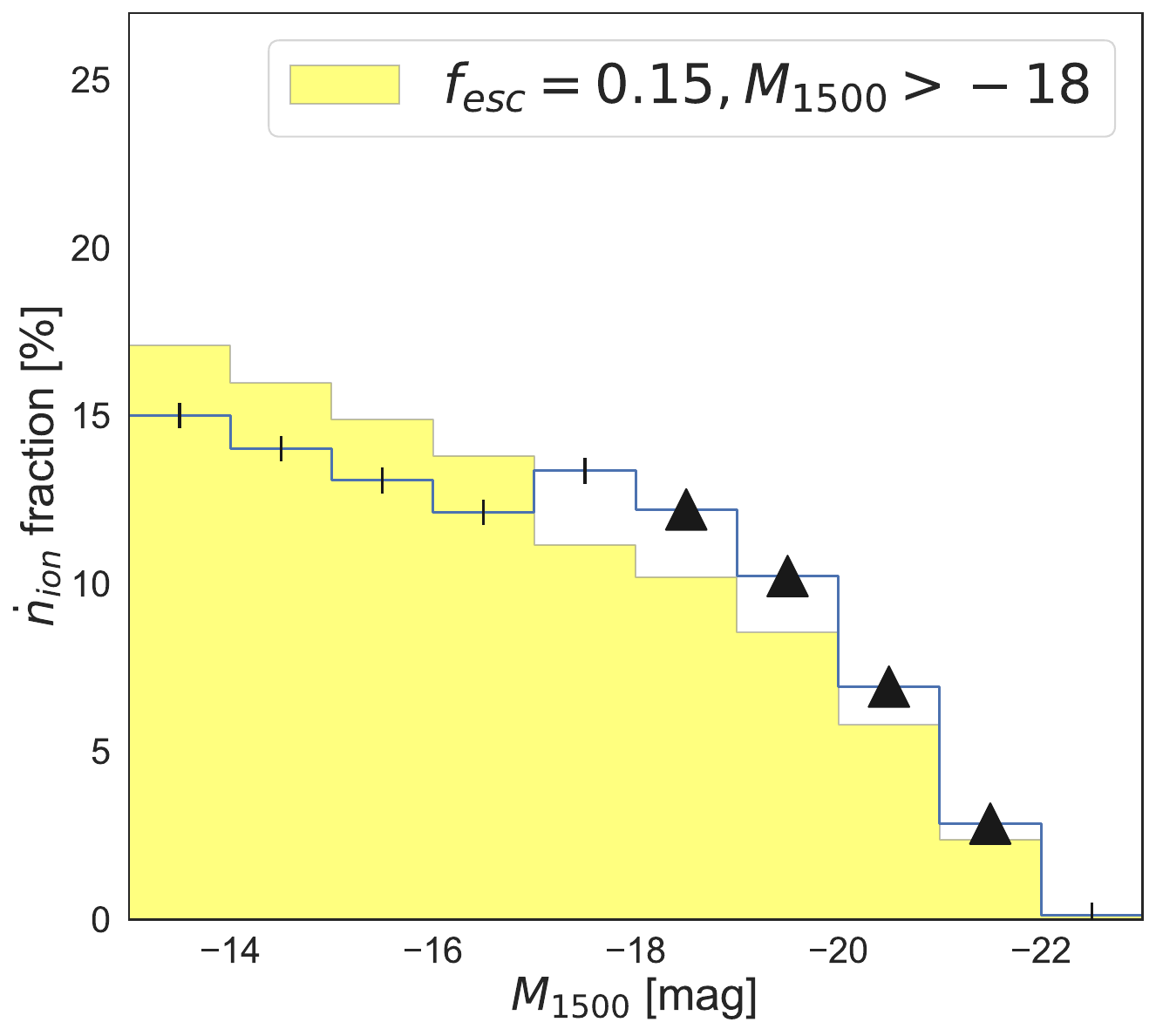}
\includegraphics[width=0.48\linewidth]{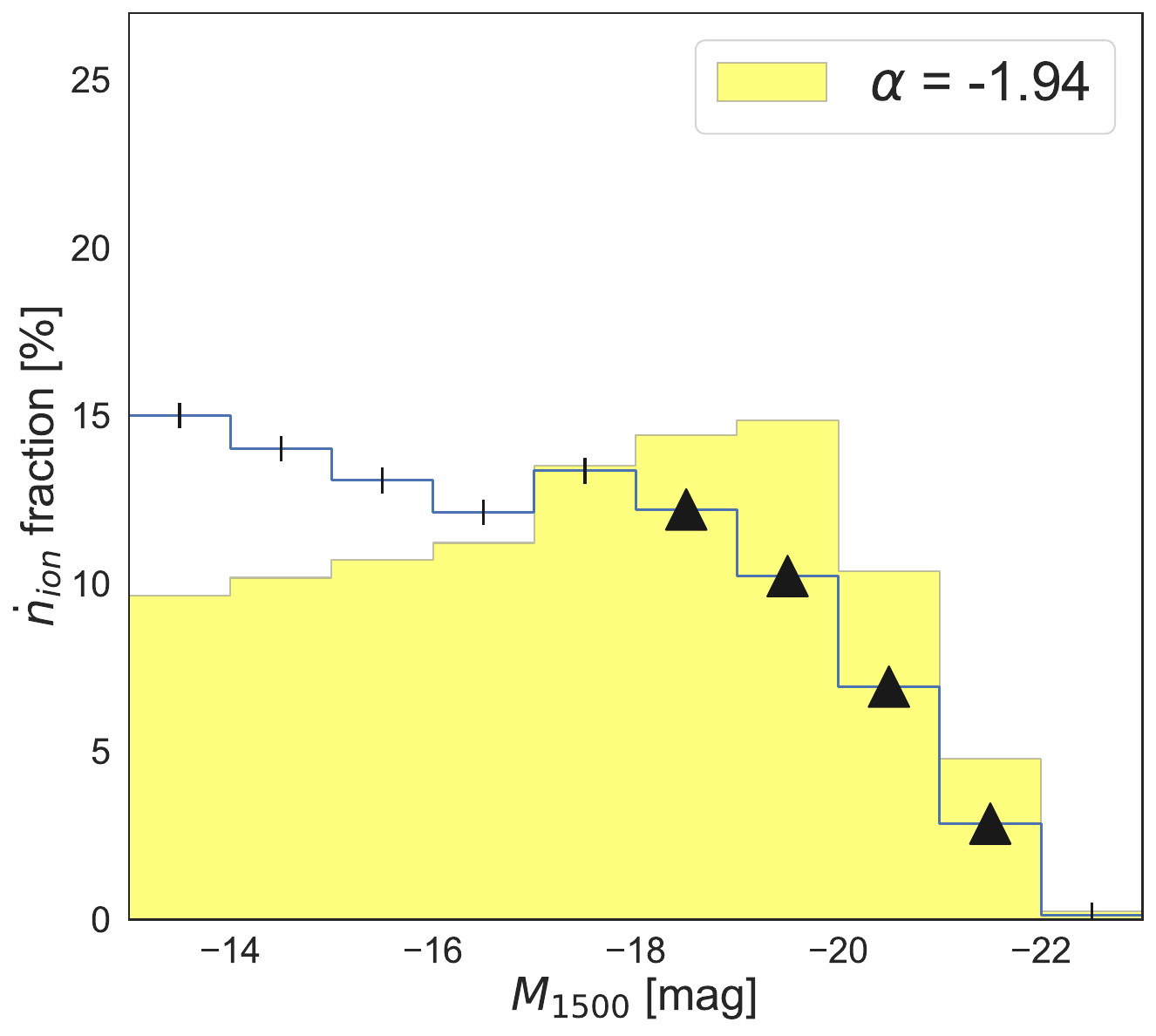}\quad
\includegraphics[width=0.48\linewidth]{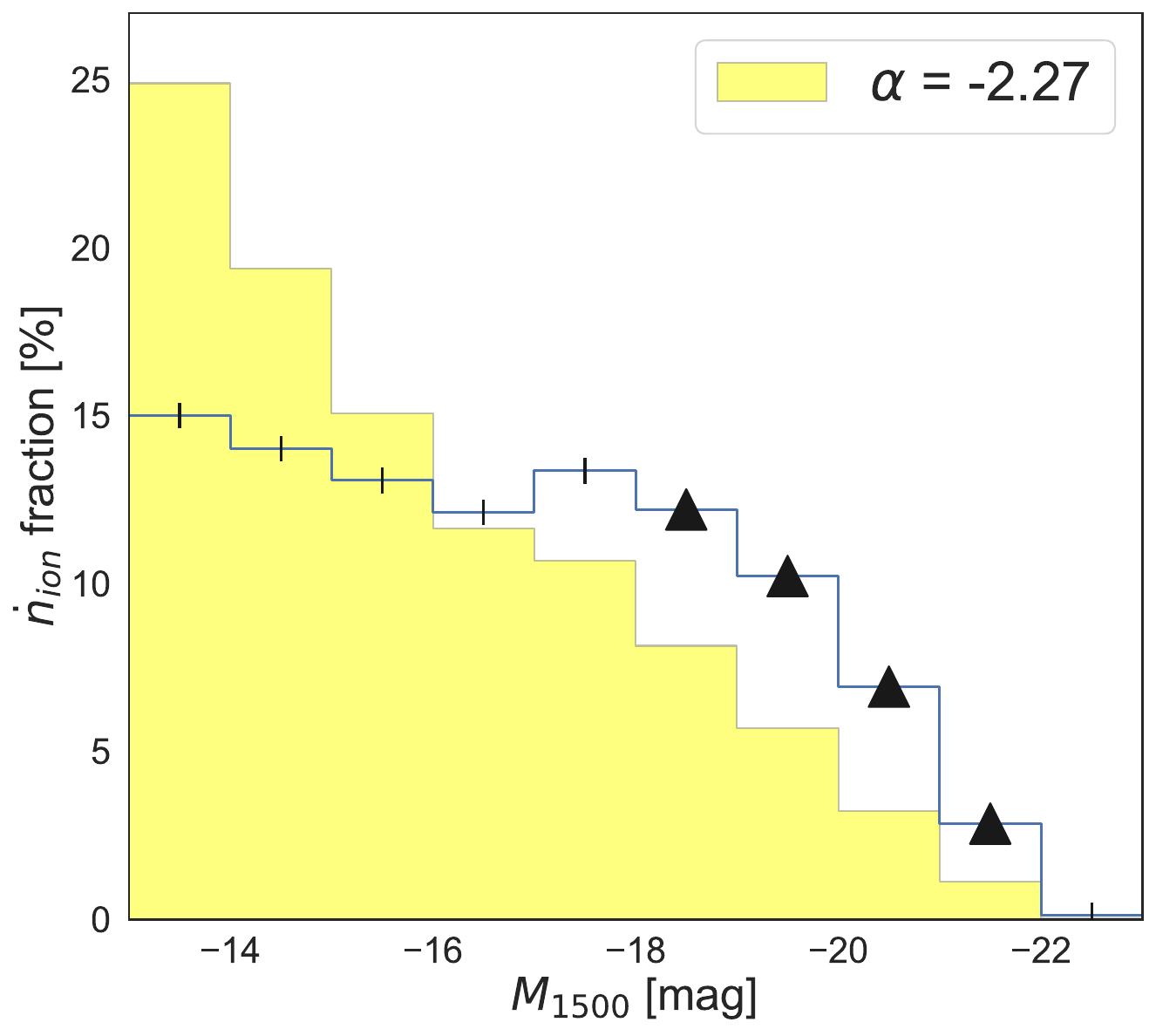}
\end{minipage}
\caption{Left: The $\dot{n}_{ion}$ fraction of galaxies at $6 \leq z \leq 9$ as function of the $M_{1500}$. Triangles represent the $\dot{n}_{ion}$ obtained from the predicted $f_{esc}$ as a function of $M_{1500}$ in each bin. Bars represent the extrapolated value, i.e. the $\dot{n}_{ion}$ fraction derived assuming that for $M_{1500}>-18$ a $f_{esc}$ value constant and equal to 0.10, and for $M_{1500}<-22$ a $f_{esc}$ value constant and equal to 0.05. 
Right: in yellow we show the $\dot{n}_{ion}$ fraction of galaxies at $6 \leq z \leq 9$ as function of $M_{1500}$ changing, respectively, from the top left to the bottom right, the $f_{esc}$ values at the faint-end (0.05 and 0.15) and the $\alpha$ parameter of the $\rho_{UV}$ ($-1.94$ and $-2.27$). The original result is also shown with the same symbols as in the figure from the left.}
\label{fig:ionizing budget}
\end{figure*}

We then derive the fraction of the total  $\dot{n}_{ion}$ that is provided by galaxies in  each magnitude bin:  the results are shown in Fig. \ref{fig:ionizing budget} and indicate that the galaxies that we can currently characterise with \jwst\ observations are contributing to only a  fraction of the total ionizing budget, i.e. less than 35\% of the total. We would therefore need to push our observations at 2-3 magnitudes deeper to characterise the bulk of the cosmic ionizers. Note however than in previous studies \citep[e.g.,][]{Finkelstein2019, Robertson2015} the fraction of ionizing photons from faint galaxies was even more prominent, with the  extreme faint end of the luminosity function dominating  the ionizing emissivity \citep[e.g., see Figure 8 of ][]{Finkelstein2019}.
We also show how the results would vary by changing the faint end $f_{esc}$ to a values of 0.05 and 0.15 respectively (right top panels of Fig. \ref{fig:ionizing budget}) and by changing the faint end slope of the UV-$\rho_{UV}$ within the 5$\sigma$ uncertainties $[-1.94, -2.27]$.
We see that the contribution of \jwst\ sources with $M_{1500}<-18$ to the integrated ionizing emissivity becomes 40\% if we assume a very small $f_{esc}$ for the faintest galaxies, or an extremely flat $LF_{UV}$ at the faint end, but it is never dominant even in these extreme cases.

\section{Summary and conclusions}\label{sec:conclusions}

In this paper, we have presented an  analysis
of 70 spectroscopically confirmed star-forming galaxies at $6 \leq z \leq 9$ from the CEERS survey, combined with 12 sources with public data from other \jwst\ early campaigns. Assuming  that the  mechanisms that facilitate the escape of LyC photons from galaxies remain consistent across all cosmic epochs, we estimated the $f_{esc}$ of the observed sources employing two empirical relations based on the most promising indirect indicators of this emission identified at $z \sim 0.3$, which are also measurable during the EoR. Using the mean inferred  $f_{esc}$  as function of $M_{UV}$ and the photon production efficiency derived from the \oiii\ emission line, we have then evaluated  the relative role of faint and  galaxies and their contribution to  the process of reionization.
Our main results  are the following:
\begin{itemize}
\item The majority of our sources show modest $f_{esc}$ values, with a mean $f_{esc}$ of  $0.13\pm0.02$, and an even lower median of  $0.08\pm0.02$.
Just 20\% of galaxies have $f_{esc}> 0.2$: 
the majority of these extreme LyC emitters show an intense O32 or high EW(\hb) coupled with small $r_e$ or very blue $\beta$ slope. As expected there is no single property that stands out as the best indirect indicator of a high LyC escape. 
\item The predicted $f_{esc}$ has a modest dependence on the total stellar mass $M_\star$ with  low mass galaxies tending to have higher mean $f_{esc}$ although the trend is scattered. 
The relation with $M_{1500}$ is less well characterised and there is not a significant dependence. 
\item There is  a strong discrepancy between our inferred  $f_{esc}$ and those predicted by  most cosmological hydrodynamical  simulations of reionization, which consistently infer much lower $f_{esc}$  values for galaxies in the same mass range as the one explored by the \jwst\ observations.
We discuss potential causes for the discrepancy such as the  failure of simulations to fully account for the bursty nature of star formation,  or the limited resolution. Alternatively using relations derived from low-redshift samples to infer $f_{esc}$ for galaxies in the EoR might not be correct and could lead to an overestimate of the $f_{esc}$ values.
\item The average predicted $f_{esc}$ have at most a modest increase with redshift from $z=5$ to $z=8$ raising from 0.11 to 0.14. 
\item The predicted $f_{esc}$ during the EoR does not show a clear dependence on the presence of \lya\ emission. This is actually expected  since in the EoR the \lya\ emission is modulated also by the IGM opacity in the local surroundings and not just by the galaxies properties as at low redshift;
\item With the inferred values of $f_{esc}$ and $\xi_{ion}$ we derive a total ionizing emissivity  $\log{\dot{n}_{ion}} = 50.50 \pm 0.38$ and $50.75 \pm 0.35\ \text{s}^{-1} \text{Mpc}^{-3}$ at redshift 8 and 6 respectively, i.e. comparable to the threshold needed to maintain the Universe ionized. Sources brighter than $M_{1500}=-18$,  which are those we can currently characterise with \jwst\ observations, only contribute less than 35\% to the total ionizing emissivity. 
\end{itemize}

The findings of this study provide crucial insights into the reionization epoch, primarily focusing on the characterization of relatively bright sources and indicating that galaxies significantly fainter and less massive than those observed by  the initial \jwst\ programs, could potentially play a dominant role in the reionization process. 
 To study significantly large samples of such faint galaxies, ultra-deep observations of galaxy cluster fields will be needed since lensing becomes a necessary tool, as in the recent work by \cite{atek2023} which reaches galaxies as faint as  $M_{UV}=-15$.
 \\In addition further work on the LyC indirect indicators will be needed to validate the fundamental assumption that the physical mechanisms and conditions that facilitate the escape of Lyman continuum photons remain the same over cosmic time. In particular, future work should be aimed at assembling a solid reference sample of Lyman continuum emitting galaxies, analogous in size to the LzLCS  survey, but at $z=3-4$, i.e. the highest redshift where a direct detection of LyC photons is possible and which is  much closer in time to the epoch of reionization. If our derived relations to infer $f_{esc}$ were still valid at z=3-4, then we could be much more  confident that they can be also applied in the EoR. 

\begin{acknowledgements}
This work is based on observations made with the
NASA/ESA/CSA James Webb Space Telescope. The
data were obtained from the Mikulski Archive for Space Telescopes at the Space Telescope Science Institute, which is operated by the Association of Universities for Research in Astronomy, Inc., under NASA contract NAS 5-03127 for JWST. These observations are associated with program JWST-ERS-01345.
We acknowledge support from the INAF Large Grant 2022 “Extragalactic Surveys with JWST” (PI: Pentericci). 
P.G.P.-G. and L.C. acknowledge support from Spanish Ministerio de Ciencia e Innovación MCIN/AEI/10.13039/501100011033 through grant PGC2018-093499-B-I00. L.C. acknowledges financial support from the Comunidad de Madrid under Atracci\'on de Talento grant 2018-T2/TIC-11612. J.S.W.L. acknowledges support from the DFG via the Heidelberg Cluster of Excellence STRUCTURES in the framework of Germany’s Excellence Strategy (grant EXC2181/1 – 390900948).

\end{acknowledgements}
\bibliographystyle{aa}
\bibliography{biblio.bib} 

\onecolumn
\section*{Appendix 1: properties of other sources}

\begin{table*}[h]
\caption{Physical and spectroscopic properties of non-CEERS galaxies}\label{tab:summary}
$$
\begin{array}{llrrccccrc}
\hline \hline
\noalign{\smallskip}
\text{PROG.} & \text{ID} & \text{RA} \ [\text{deg}] & \text{DEC} \ [\text{deg}] & z_{spec} & \beta &EW_0(\hb) \ [\AA]& O32 & r_e \ [\text{kpc}] & f_{esc}\ \text{(pred.)}\\
\noalign{\smallskip}
 \hline
 \noalign{\smallskip}
\textbf{DDT} &
 10025 & 3.59609 & -30.38581 & 7.875 & -2.08 \pm 0.45 & 139 \pm 30& 6.6 \pm 1.4 & 0.40 & 0.11 \pm 0.09\\
& 100004 & 3.60657 & -30.38093 & 7.884 & -1.88 \pm 0.44& > 130 & > 5 & 0.40 & 0.07 \pm 0.05\\
\noalign{\smallskip}
\hline
\noalign{\smallskip}
\textbf{GLASS} &
10000 & 3.60134 & -30.37923 & 7.884 & -2.27 \pm 0.46 & 76 \pm 13 & 8 \pm 3 & 0.20 & 0.21 \pm 0.16\\
& 10021 & 3.60851 & -30.41854 & 7.288 & -2.25 \pm 0.48 & 104 \pm 24 & 13 \pm 5 & 0.68 & 0.10 \pm 0.09\\
& 100001 & 3.60385 & -30.38223 & 7.875 & -1.63 \pm 0.48 & 39 \pm 8 & 3.2 \pm 1& 0.50 & 0.04 \pm 0.03\\
& 100003 & 3.60451 & -30.38044 & 7.880& -2.51 \pm 0.48 & 85 \pm 18 & 21 \pm 7 & 0.15 & 0.40 \pm 0.22\\
& 100005 & 3.60646 & -30.38099 & 7.883& -2.55 \pm 0.48 & 33 \pm 15 & 2.9 \pm 1& 0.25 & 0.15 \pm 0.12\\
& 150008 &3.60253& -30.41923& 6.230 & -2.10 \pm 0.25 & 141 \pm 30 & > 20 & 0.40 & 0.08 \pm 0.07\\
& 400009 &3.60059 &-30.41027& 6.376 & -2.17 \pm 0.25 & 35 \pm 7 &- & 0.11 & 0.07 \pm 0.06\\
\noalign{\smallskip}
\hline
\noalign{\smallskip}
 \textbf{ERO} &4590 & 110.8593287& -73.4491656& 8.496& -2.20 \pm 0.15 & 218 \pm 150& >14.8 & 0.71 & 0.08 \pm 0.07\\
&5144 & 110.8396739& -73.4453570& 6.378 & - & 151 \pm 51 & 18.6 \pm 3.3 & 0.92 & -\\
&6355 &110.8445942 & -73.4350590& 7.665 & -1.96 \pm 0.22 & 150 \pm 4 & 8.2 \pm 0.3 &0.83 & 0.03 \pm 0.02\\
&10612 & 110.8339649& -73.4345232& 7.660 & -2.31 \pm 0.11 & 210 \pm 16& 14.8 \pm 1.7 & 0.42 & 0.14 \pm 0.12\\
\hline
\noalign{\smallskip}
\end{array}
$$
\end{table*}

\section*{Appendix 2: an analysis on the empirical relation calibrated on the LzLCS sample}

\begin{figure}[ht!]
\includegraphics[width=0.5\linewidth]{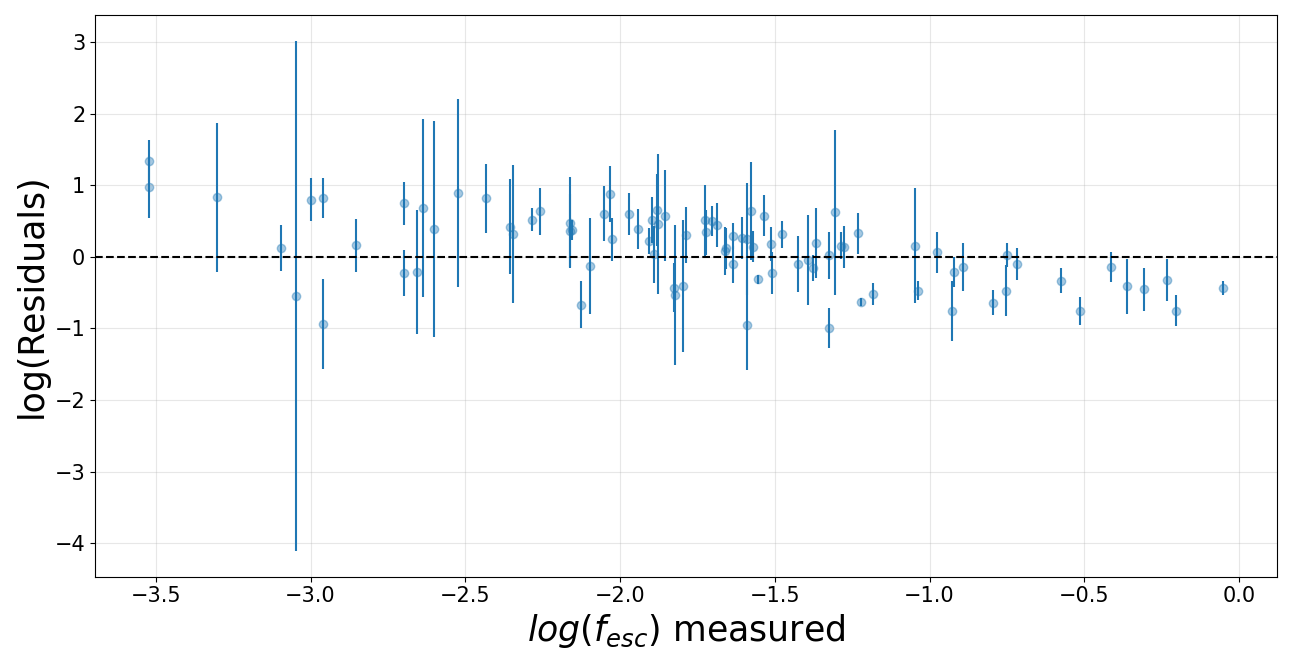}
\includegraphics[width=0.5\linewidth]{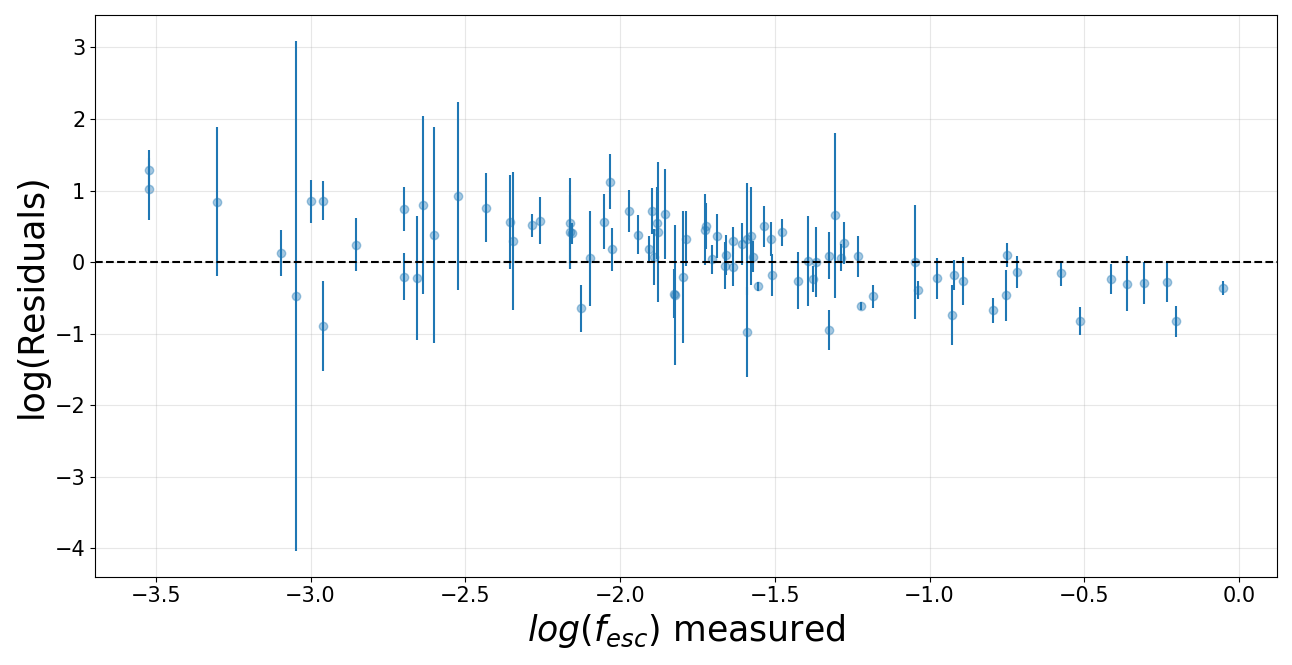}
\caption{Left: The difference between the measured values and the values estimated by the  M23 empirical relation. Right: Same for the values estimated using for Eq. \eqref{eq} of this paper. }
\label{fig:residuals}
\end{figure}

To accurately assess the reliability of the empirical relation calibrated on the LzLCS presented in two versions (Eq. 1 in M23 and Eq \eqref{eq} of this study) for estimating the $f_{esc}$ values using indirect indicators, we examined the distribution of the residuals which are  defined as the difference between the measured values and the values estimated by our relation. In Figure \ref{fig:residuals}, we plot the residuals obtained using both versions of our relation plotted against the real values, in logarithmic scale. From the plots, it is evident that the two relations exhibit statistical equivalence, as their residual distributions are identical. we can also see that our relation tends to underestimate the $f_{esc}$ values for $f_{esc} >0.1$ and overestimate them for $f_{esc}$ below 0.01. This outcome is a direct consequence of the initial sample, which is predominantly composed of sources with modest $f_{esc}$, around 0.02-0.05.

\end{document}